\documentclass[useAMS,usenatbib]{mn2e}

\usepackage{aas_macros} % Enable AAS-style references from ADS
\usepackage{graphicx} % For \includegraphics
\usepackage{tabularx} % Automatically calculate table width
\usepackage{multirow} % For multi-row tables used in source list
\usepackage{rotating}
\usepackage{hhline} % Draws double horizontal lines in tables
\usepackage{subfigure} % Use subfigures 
\usepackage{amssymb} % AMS symbols like \gtrsim
\usepackage{flafter} % Forces floats to appear after the defining text
\usepackage{placeins} % Forces floats to before \FloatBarrier statement
\usepackage{bm} % boldmath?
\usepackage{mdwlist} % for suspend/resume lists
\usepackage{psfrag}
\usepackage{amsmath} % for align env
\usepackage{chngpage} % for too-wide tables
\usepackage{subfloat} % for splitting figures over > 1 page

\usepackage[T1]{fontenc}
\usepackage{aecompl}
\newcommand{\Ks}{\ensuremath{K_{\mathrm{s}}\,}}

\usepackage{color} % Add coloured text
\newcommand{\referee}[1]{#1}
\newcommand{\updated}[1]{#1}
\newcommand\T{\rule{0pt}{2.6ex}}
\newcommand\B{\rule[-1.2ex]{0pt}{0pt}}

\title[Star-formation history of VIDEO-selected galaxies]{The
  star-formation history of mass-selected galaxies from the VIDEO
  survey}

\author[\updated{J.~T.~L.~}Zwart et al.]{
Jonathan T.~L.~Zwart\thanks{email: jzwart@uwc.ac.za},$^{1}$
Matt J.~Jarvis,$^{1,2}$
Roger P.~Deane,$^{3}$
David G.~Bonfield,$^{4}$
\newauthor
Kenda Knowles,$^{5}$
Nikhita Madhanpall,$^{1,6}$
Hadi Rahmani$^{7,8}$ and
Daniel J.~B.~Smith$^{4}$\\
$^1$ Department of Physics \& Astronomy, University of the Western Cape, Private Bag X17, Bellville 7535, South Africa\\
$^2$Astrophysics, Department of Physics, Keble Road, Oxford OX1 3RH\\
$^3$Astrophysics, Cosmology \& Gravity Centre, Astronomy
Department, University of Cape Town, Private Bag X3, Rondebosch 7701, South Africa\\
$^4$Centre for Astrophysics, Science \& Technology Research Institute, University of Hertfordshire, Hatfield, Hertfordshire AL10 9AB\\
$^5$Astrophysics and Cosmology Research Unit, School of Mathematical Sciences, University of KwaZulu-Natal, Durban, 4041, South Africa\\
$^6$School of Physics, University of the Witwatersrand, 1 Jan Smuts Avenue, Braamfontein, Johannesburg, 2050 South Africa\\
$^7$Inter-University Centre for Astronomy and Astrophysics, Post Bag 4, Ganeshkhind, Pune 411\,007, India\\
$^8$School of Astronomy, Institute for Research in Fundamental Sciences, PO Box 19395--5531, Tehran, Iran
}

\begin{document}

\date{Accepted ---. Received ---; in original form \today.}

\pagerange{\pageref{firstpage}--\pageref{lastpage}}
\pubyear{2013}

\label{firstpage}

\maketitle

%------------------------------------------------------

\begin{abstract}
  \referee{W}e measure \updated{star-formation rates (SFRs)} and
  specific SFRs (SSFRs) \updated{of} \Ks-selected galaxies from the
  VIDEO survey \updated{by stacking 1.4-GHz Very Large Array data.}
  \referee{We split the sample, which spans $0<z<3$ and
    stellar masses $10^{8.0}<M_*/M_{\odot} <10^{11.5}$, into
    elliptical, irregular or starburst galaxies based on their
    spectral-energy distributions.} We find that SSFR falls with stellar mass, in
  agreement with the `downsizing' paradigm. We consid\referee{er} the
  dependence of the SSFR--mass slope on redshift: for our full and
  elliptical samples \updated{the} slope flattens, but for the
  irregular and
  starburst samples the slope is independent of redshift. \referee{The
    rate of SSFR evolution reduces} slightly with stellar mass for
  ellipticals, but irregulars and starbursts co-evolve \referee{across
    stellar masses}.

  Our results for SSFR as a function of stellar mass and redshift are
  in agreement with those derived from other radio-stacking
  measurements of mass-selected passive and star-forming galaxies, but
  inconsistent with those generated from semi-analytic models, which
  tend to underestimate SFRs and SSFRs. There is a need for deeper
  high-resolution radio surveys such as those from telescopes like the
  next-generation MeerKAT in order to probe lower masses at earlier
  times and to permit direct detections, i.e.~to study individual
  galaxies in detail.
\end{abstract}


%------------------------------------------------------
\section{Introduction}
\label{sec:intro}

Untangling the star-formation history of galaxies is of basic
importance in validating our knowledge of cosmology via astrophysics;
it permits measurement of the build-up of galactic stellar mass,
provides constraints on initial conditions, pins down supernova rates
and allows us to compare models of chemical evolution (see
e.g.~\citealt{hopkins2006}, \citealt{sfrcompare2012}). The
star-formation rate density of the Universe \citep{madau1998} is
well-constrained to $z\approx 2$ but at higher redshift ($2<z<5$)
there is still some dispute using different wavelengths. Crucial for
calculating the star-formation rate density at higher redshifts, the
star-formation rate (SFR) is often determined in one of the following
ways (\citealt{calzetti2012}, \citealt{kennicutt2012} and
\citealt{sfrcompare2012} give excellent reviews):

\begin{enumerate}

\item UV observations (e.g.~\citealt{lilly1996},
  \citealt{steidel1999}, \citealt{wilson2002}, \citealt{feulner2007},
  \citealt{elbaz2007}, \citealt{zheng2007}, \citealt{damen2009},
  \citealt{ellis2013}) reach relatively high redshifts ($z\gtrsim6$),
  but are severely obscured by dust, and hence miss a large fraction
  of the star formation. Emission is from massive, short-lived
  stars. The SFR is sometimes calculated using e.g.~the
  \cite{bell2005} relation together with the infra-red luminosity
  $L_{\mathrm{IR}}$, hence $L_{\mathrm{UV+IR}}$, or simply directly
  from the UV \citep{hilton2012}. UV-derived SFRs have been found to
  be systematically lower than IR-derived SFRs (see
  e.g.~\citealt{hilton2012} and Figure 4 of \citealt{burgarella2013}),
  and UV-derived specific SFRs (SSFRs) show a relative deficit at
  higher masses compared to radio-derived SSFRs
  (\citealt{pannella2009}; see also below)\updated{.}

\item In the rest-frame optical, star-formation rates can be
  determined from recombination lines such as H$\alpha$, H$\beta$ and
  [O\textsc{ii}], with photons emitted by nebulae around young,
  massive (OB) stars (see e.g.~\citealt{kk1983},
  \citealt{moustakas2006}). However, emission lines must also be
  corrected for attenuation by dust and variations in excitation, and
  studies are restricted to low redshifts ($z\lesssim 0.5$) and
  limited precision, unless use is made of near-infrared spectroscopy
  (\citealt{roseboom2012a}, \citealt{roseboom2013a}). Yet \cite{ds2012}
  find agreement between emission-line indicators and IR indicators
  (see next point)\updated{,} as well as concluding that metallicity
  plays an important role.

\item Submm/FIR emission is generated from UV photons re-emitted by
  dust grains in star-formation regions and reveals star formation
  that is otherwise optically obscured. One can constrain SFRs by
  estimating $L_{\mathrm{IR}}$ (8--1000\,$\mu$m) via spectral-energy
  distribution (SED) fitting, then translating that quantity to
  \updated{a} SFR
  via the \cite{kennicutt1998b} relation (e.g.~\citealt{polletta2008},
  \citealt{cava2010}). However, extrapolations to higher redshift
  and/or lower SSFR must be made because (a) relatively poor telescope
  resolution for single-dish, space-based telescopes such as
  \textit{Herschel} leads to rapid source confusion and (b) observing
  from the ground requires excellent weather. There are also two
  competing systematic effects that tend to under- and over-estimate
  the SFR respectively: some starlight is \textit{not} absorbed by
  dust, and evolved stars contribute to dust heating, something that
  may only be accounted for with full SED modelling
  (e.g.~\citealt{magphys2012}, \citealt{dan2012}). In spite of all
  this, recent results indicate that the cosmic star-formation rate
  density levels off at $z\approx 3$ then begins to drop at higher
  redshifts (e.g.~\citealt{hopkins2006}, \citealt{lapi2011},
  \citealt{burgarella2013}\referee{, \citealt{behroozi2013b}}).

\item Deep radio surveys are able to probe the galaxy SFR because of
  cosmic-ray and synchroton emission from accelerated electrons in the
  magnetic fields of supernova remnants \citep{helou1985}. The
  relationship between SFR and 1.4-GHz luminosity is calibrated to the
  far-infrared-radio correlation (e.g.~\citealt{condon1992},
  \citealt{haarsma2000}, \citealt{yun2001}, \citealt{condon2002},
  \citealt{bell2003}). Radio-wavelength observations are not obscured
  by dust and their higher angular resolution drastically reduces
  source confusion, but they may also suffer from AGN contamination
  (as is the case for all other tracers) \updated{and traditionally do
  not reach deep-enough flux densities to make them useful for
  studying star formation at high redshift.}

\item X-rays are linked to the SFRs of late-type galaxies because of
  neutron-star X-ray binaries, and supernova remnants, which therefore
  couple to young stellar populations having recent star formation,
  but they may also suffer from AGN contamination. See
  \cite{norman2004} and \cite{daddi2007b} for applications of this
  method. \cite{zinn2012} have since measured SFRs and SSFRs in
  stacking \textit{Chandra} data and found much lower rates than in
  the radio, although the trend of increasing SSFR out to $z=3$ is at
  least consistent between the two.

\end{enumerate}

\noindent In general, estimates of \updated{SFRs} from these diverse
routes can differ (see e.g.~\citealt{sfrcompare2012}) because of
selection criteria, calibrations in the different wavebands, the
assumed initial mass function (IMF) and SEDs. By way of an example,
SFRs derived from 24-$\mu$m fluxes via the infra-red luminosity can be
overestimated because of (dominant) contamination by evolved stars
\citep{kennicutt2012} or by AGN contamination (see
e.g.~\citealt{alejo2005}, \citealt{alejo2006}).

The SSFR is a measure of a galaxy's star-formation efficiency,
i.e.~the fraction of its mass that could be transformed into stars at
a given cosmic time \citep{dunne2009}. In particular, it informs us
about the phenomenon of `downsizing' \citep{cowie1996,pg2008}, whereby
the most massive galaxies have the lowest SSFRs at all redshifts (see
e.g.~\citealt{schim2007}), and so formed their stars earlier and more
rapidly than those of lower mass \citep{rodighiero2010}. In other
words, the dominant star-forming population has slowly moved to
lower-mass galaxies over cosmic time \citep{kennicutt2012}. Although
downsizing is widely accepted as the prevailing explanation for the
slope of SSFR with stellar mass, a key aim of current and future
surveys is to quantify it.

The recent wave of deep near-infra-red surveys, such as the United
Kingdom Infrared Deep Sky Survey Ultra-Deep Survey (UKIDSS--UDS;
\citealt{ukidss2007}) and UltraVISTA \citep{ultravista} are able to
constrain stellar mass\updated{. The} photospheric-emission peak of
evolved stars is at roughly 1--2\,$\mu$m, \updated{therefore}
selection in the \Ks band is a good approximation to selection by
stellar mass (see e.g.~\citealt{daddi2007}). However, if the SFR is to
be calculated from radio emission, counterpart surveys tend to be
shallower than in the near infra-red, so it has become commonplace to
bin objects statistically (so-called `stacking'; see
e.g.~\citealt{serjeant2004}, \citealt{dole2006}, \citealt{ivison2007},
\citealt{takagi2007}, \citealt{white2007}, \citealt{papovich2007},
\citealt{dunne2009}, \citealt{garn2009}, \citealt{magdis2010},
\citealt{bourne2011}, \citealt{karim2011}\referee{,
  \citealt{heinis2013},
  \citealt{viero2013}}).

Employing stacking in the radio, \cite{dunne2009} and \cite{karim2011}
selected populations of normal and star-forming galaxies in the K-band
using a colour-colour method (BzK) and SED fitting respectively.
\updated{\citet{dunne2009} investigated the star-formation history of
  BzK-selected galaxies from UKIDSS--UDS using stacking of 610-MHz and
  1.4-GHz data from the VLA and the Giant Metre-wave Telescope (GMRT)
  respectively. In the analysis of data from the Cosmic Evolution
  Survey (COSMOS) field, \cite{karim2011} selected galaxies at
  $3.6\mu$m, stacking 1.4-GHz VLA data (A and C arrays), with a noise
  of $8\mu$Jy at the centre of their 1.72\,deg$^{2}$\,map. They
  calculated stellar masses using SED fitting from their
  photometric-redshift \updated{fitting.}} There was good agreement in
SSFR--redshift evolution between these studies, but the dependence of
SSFR on stellar mass was found to be much shallower for the UKIDSS
data than for COSMOS. \cite{karim2011} attribute the difference to
discrepant stellar-mass estimation \updated{methods}, rather than to
e.g.~cosmic variance: \cite{dunne2009} calculated stellar masses via
the absolute K-band magnitude, which might transfer low-mass (where
the conversion is less applicable) star-forming galaxies with high
SSFRs to higher masses, causing the SSFR--$M_*$ slope to flatten
\citep{karim2011}.

In this work we present an independent \referee{`}stacking\referee{'}
analysis at 1.4\,GHz for sources from the VISTA Deep Extragalactic
Observations (VIDEO) survey \citep{jarvis2013}, in order to measure
star-formation rates of mass-selected galaxies. \referee{The available
  10-band photometry allows \updated{accurate} SED
 photometric-redshift and stellar-mass estimation, \textit{and}
 classification of sources into elliptical, irregular \updated{and
   starburst.} At the same time, there is an opportunity to test SSFRs
 determined from semi-analytic models (e.g.~\citealt{henriques2012})
 against observations of a mass-selected sample}.

In section \ref{sec:obs} we outline the data in hand, going on to
describe photometric-redshift and stellar-mass estimation, and the
sample selection, in sections \ref{sec:seds} \updated{and}
\ref{sec:seln}. Our stacking method is presented in section
\ref{sec:stacking}, and we \referee{subsequently} describe how we
calculate SFRs and SSFRs from radio fluxes/luminosities. Our results
and discussion appear in sections \ref{sec:results} and
\ref{sec:discussion} and we conclude in section \ref{sec:conclusions}.

We assume radio spectral indices $\alpha$ are such that
$S_{\nu}\propto\nu^{\alpha}$ for a source of flux density $S_{\nu}$ at
frequency $\nu$. All coordinates are epoch J2000. Magnitudes are in
the AB system. We assume a $\Lambda$CDM `concordance' cosmology
throughout, with $\Omega_{\mathrm{m}}=0.27$, $\Omega_{\Lambda}=0.73$
and $H_0= 70\,\mathrm{km\,s^{-1}}\,\mathrm{Mpc}^{-1}$
\citep{wmap7params}.

\section{Observations}
\label{sec:obs}

\subsection{Infrared and Optical Observations}
\label{sec:obs:nir}

The \referee{ongoing} VIDEO Survey will \referee{eventually} cover
12~square degrees over three fields with a view to studying galaxy
formation, evolution and clusters. The data presented here, covering
$\approx$ 1~square degree, were taken in $Z, Y, J, H,$ and \Ks
bands. We also use optical data from the overlapping
Canada--France--Hawaii Telescope Legacy Survey (CFHTLS--D1;
\citealt{ilbert2006}) in the $u^*, g', r', i', $ and $z'$ bands. The
5-$\sigma$, 2-arcsec-diameter aperture AB-magnitude limits are, as of
June 2012, $u$*=28.7, $g$\arcmin=28.8, $r$\arcmin=28.4,
$i$\arcmin=28.1, $z$\arcmin=27.2, $Z$=25.7, $Y$=24.6, $J$=24.5,
$H$=24.0 and \Ks=23.8. Full details can be found in \cite{jarvis2013}.

\subsection{Radio Observations}
\label{sec:obs:radio}

The radio data are described by \citet{bondi2003}, so we give only a
summary. The 1.4-GHz Very Large Array (VLA; B-array) observations of
\cite{bondi2003} cover a 1-square-degree field centred on
J\,2$^h$\,26$^m$\,00$^s$ \,$-$4$\degr$\,30$\arcmin$\,00$\arcsec$
(i.e.~the XMM--LSS field). A mosaic of nine pointings yields a
variation in noise of about 20 per cent around $17.5\,\mu$Jy. The FWHM
of the \textsc{clean} restoring beam is 6\,arcsec and the map has 2048
$\times$ 2048 1.5-arcsec pixels. \textsc{clean} bias and bandwidth
smearing \referee{were} two issues considered and resolved in the analysis of
\cite{bondi2003}.
\referee{In our stacking analysis confusion from sources below the
  flux limit may bias our results. However, with a 6\,arcsec
  synthesized beam we use equation 27 from \cite{condon2012} to
  estimate the expected source confusion given this beam. We calculate
a confusion noise of 0.8$\,\mu$Jy\,beam$^{-1}$, which is significantly
fainter than the signal we measure from the stacking analysis in
section \ref{sec:stacking}.}

\section{Spectral-Energy Distributions}
\label{sec:seds}

\subsection{Photometric-Redshift Estimation}
\label{sec:seds:zphots}

\begin{figure}
\centering
\includegraphics[width=8cm,origin=br,angle=0]{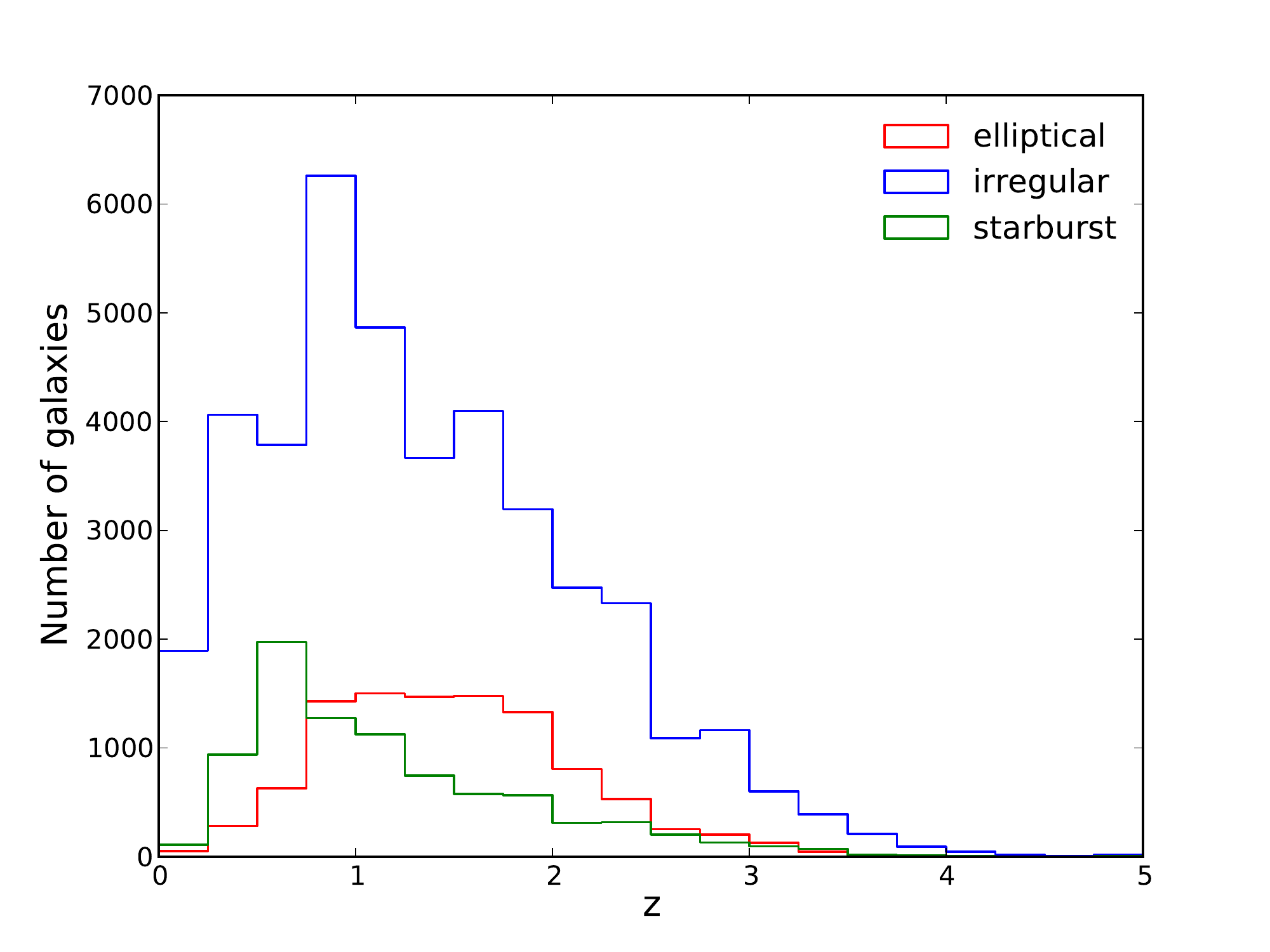}
\caption{Histogram of best-fit photometric redshifts for each galaxy
  sample after the data selection described in section
  \ref{sec:seln}. \updated{Although in this work we only consider
    sources for which $z<3$, the histogram shows that there are a
    number of sources at redshifts $z>4$.}\label{fig:zhistos}}
\end{figure}

Photometric redshifts were determined using \textsc{Le Phare}
(\citealt{arnouts1999}, \citealt{ilbert2006}, \citealt{ilbert2009}).
SED templates are taken from \cite{cww1980} with a starburst SED due
to \cite{kinney1996}, plus a series of SEDs interpolated between all
of these and extended with the GISSEL synthetic models from
\cite{bc03}. \cite{jarvis2013} provide more details. A histogram of
the best-fit photometric redshifts is shown in Figure
\ref{fig:zhistos}, where we note that our \Ks-band selection is
sensitive to the older stellar populations.

\cite{bauer2011} helpfully simulated the effect of using photometric
redshifts in the calculation of SFRs and SSFRs, including catastrophic
outliers. The assumed photometric redshift uncertainties were gaussian
with $\Delta z/(1+z)\approx0.1$ over the range $1.5<z<3$. They found
that on average their SFR and SSFR relations were robust even with
these uncertainties included, though the uncertainty did increase for
less-massive galaxies.
\referee{To mitigate the effect of uncertain
  photometric redshifts, we
  employ relatively large redshift bins compared to the typical
  photometric uncertainty of $\Delta z/(1+z)$=0.13 for the VIDEO
  survey \citep{jarvis2013}. However in section
  \ref{sec:results:comparison:karim11} we simulate the effects that
  photometric uncertainties have on our results.}

\subsection{Stellar-Mass Estimation}
\label{sec:seds:masses}

With photometric-redshift estimates fixed at the best-fit values found
using the SED templates, we fitted \textsc{galaxev} templates to each
object's photometry using \textsc{Le Phare} in order to obtain an
estimate of stellar mass $M_*$ (following \citealt{ilbert2009}). We
corrected masses obtained from 2-arcsec aperture photometry to total
masses using an empirically-derived, redshift-dependent aperture
correction.

The templates we considered had three different metallicities
($Z_{\odot}=$ 0.004, 0.008 and 0.02) and nine different
exponentially-decreasing star-formation rates
$\propto\mathrm{e}^{-\tau}$, with $\tau=$ 0.1, 0.3, 1, 2, 3, 5, 10, 15
and 30\,Gyr. We allowed for a dust-extinction correction with
$0<E\left(B-V\right)<0.5$. In each case we assumed a Chabrier initial
mass function \citep{chabrier2003} and that ages are less than that of
the Universe at the corresponding redshift.

\referee{The median logarithmic stellar-mass uncertainty for the full
  sample is $\log_{10}\left(\Delta M_*/M_{\odot}\right)=0.11$. These
  uncertainties of course depend on photometric-redshift
  uncertainties. However, we note that our binning scheme of 0.5 in
  $\log_{10}\left(M_*/M_{\odot}\right)$ means that this does not
  influence our results.}

\section{Data Selection}
\label{sec:seln}

\subsection{Galaxy Classification}
\label{sec:seln:sample}

We study several different types of population in this work, all
selected at \Ks band. \referee{Galaxies can be classified using a
  colour-colour diagram (for example, $BzK$; see
  e.g.~\citealt{dunne2009}). However, w}e classif\referee{ied}
galaxies based on the best-fitting templates described in section
\ref{sec:seds:zphots}, in order to investigate the differences between
passive and star-forming galaxies. We divide\referee{d} the templates
into three sets: ellipticals; Sbc, Scd and irregular galaxies; and
starburst galaxies. \referee{Template-based classification takes into
  account
  the full SED (via 10-band photometry) and ought to be a better
  discriminator than a colour-colour diagram. Although there may be
  overlaps, we have split into very broad types using all the
  available information and as such these types are likely to be
  robust.} We then cut the data in various ways in order to remove
contaminants, as detailed in the following sections.

\subsection{Star-Galaxy Separation}
\label{sec:seln:cuts:stars}

Following \cite{jarvis2013}, we classify objects as stars if

\begin{equation}
J-K<0.3+f_{\mathrm{locus}}\left(g-i\right)
\end{equation}

\noindent where

\begin{align}
f_{\mathrm{locus}}(x) & = & - & 0.7127 & & & x<0.3 \notag \\
f_{\mathrm{locus}}(x) & = & - & 0.89+0.615x-0.13x^2 & \mathrm{for}\hspace{1pt} & & 0.3<x<2.3 \notag \\
f_{\mathrm{locus}}(x) & = & - & 0.1632 & & & x<2.3 \notag, \\
\end{align}

\noindent and where $x=g-i$ and the global offset of 0.1\,mag from the
\cite{baldry2010} relation accounts for the difference between
K$_{\mathrm{s}}$ and K. We remove\referee{d} all sources that
reside\referee{d} in the area of this colour space occupied by stellar
sources.

\subsection{Further Selection Considerations}
\label{sec:seln:cuts}

\referee{Having classified objects by galaxy type, or as stars, we
  carried out further data selections as follows:}

\begin{enumerate}
\item We only consider objects for which \Ks$<$23.5 to ensure that we
  have a `complete' sample, since the formal 5-$\sigma$ limit is 23.8
  in \Ks band; the completeness of VIDEO at \Ks=23.5 was found by
  \cite{jarvis2013} to be $>90$~per~cent.
\item We restrict our analysis to objects for which the best-fit
  redshift $z<3.0$, since the VIDEO data reach $L^*$ by $z\simeq 3.5$;
  the VLA data are also insufficiently deep to probe to higher
  redshifts.
\item Some regions of the VIDEO field are contamined by ghosting
  haloes coming from detector-reflected star light; we therefore
  excise any \Ks-band objects that lie close to the very brightest
  stars.
\item In several cases templates are not (or are poorly) fitted by
  \textsc{Le Phare} because the data do not sufficiently constrain any
  model. We only use sources with relatively good photometric-redshift
  estimates, and therefore impose a template goodness-of-fit
  $\chi^2<50$. This will not bias our results as we only discuss
  evolutionary trends per source class, and do not tackle the issue of
  evolution per volume density where such a selection would need to be
  accounted for.
\item We remove objects with non-zero \textsc{SExtractor} flags,
  i.e.~those that are too crowded, blended, saturated, truncated or
  otherwise corrupted.
\end{enumerate}

Of course the fractions removed by each cut depend on the order in
which the cuts are applied, and some objects will be multiply-cut;
these fractions should therefore only be considered to be approximate:
from the original catalogue of \Ks-selected objects,
\updated{14.0}~per~cent fail the redshift cut, \updated{3.9}~per~cent
are removed because of haloes, \updated{16.3}~per~cent have non-zero
\textsc{SExtractor} flags and \updated{8.5}~per~cent have
poorly-fitting \textsc{Le Phare} SED templates.

Table~\ref{table:species} summarizes each sample. \referee{The final
  redshift and stellar mass ranges are $0<z<3$ and
  $10^{8.0}<M_*/M_{\odot} <10^{11.5}$ respectively.} As one moves from
elliptical towards starburst galaxies, the median redshift $<z>$
decreases because of the generally lower masses of late types as
compared to early types, combined with our \Ks magnitude limit.

\section{Stacking}
\label{sec:stacking}

\referee{The term `stacking' is vernacular and thus care must be
  taken over its use. \cite{blast2009} defined it as taking the
  covariance of a map with a catalogue, while other authors use some
  variation on the technique while at the same time not
  attempting a definition.
  \textit{We} take `stacking' to mean} the process of combining data
from one set of observations (in this case, the VLA data) using the
positions of sources selected from another data set (here, the VIDEO
catalogue). The aim is to increase the signal-to-noise ratio for a
particular galaxy sample and so describe that population in a
statistical way. \referee{The combining operation is defined below.}

\referee{There are many techniques in the literature for stacking data
  sets (see the examples given in section \ref{sec:intro}), but two
  that might be applied here are (i) stacking pixels and
  assessing the distributions via a \updated{mean or median} (see
  e.g.~\citealt{dunne2009}, \citealt{karim2011}), or (ii) modelling
  the histograms of flux distributions
  (e.g.~\citealt{ketron2013}). Here we consider the first, while the
  second will be presented in a forthcoming paper.}

In pixel stacking (see e.g.~\citealt{white2007}), one combines pixel
values in the radio-flux map wherever there is a known \Ks-selected
source (\citealt{dunne2009}, \citealt{garn2009}), to give a
distribution of fluxes for that particular class of object. Since each
\Ks-selected source has a corresponding estimate of a photometric
redshift and stellar mass, it is further possible to calculate, from
these and the radio flux, the radio luminosity of sources as well as
the SFRs and SSFRs (see section \ref{sec:sfrs}). One therefore
obtains---for each sample---distributions of fluxes, luminosities,
SFRs and SSFRs.

The flux and other distributions are in general non-gaussian and
asymmetric \updated{because of bright/detected sources and the
  underlying
  source-count distribution}, so it is not appropriate to simply
describe them using a mean and its standard error. \referee{On the
  other hand, a median can, under some circumstances, be subject to
  its own biases. For example, in their Appendix, \cite{bourne2012}
  identify three potential sources of such bias: the flux limits and
  underlying shape of the `true' distribution\updated{,} and the
  magnitude of the
  thermal noise. Although \cite{bourne2012} \textit{did} detect a bias
  in simulations, it remained small compared to true variations in
  stacked fluxes, and---crucially---all trends remained significant
  and conclusions unaffected whether or not a bias correction was
  applied.}

\referee{We opt for the median
  flux (or redshift, luminosity, SFR, stellar mass or
  SSFR) in this analysis because of
  its clearly reduced sensitivity to high-flux (or redshift,
  luminosity, SFR, stellar mass or SSFR) outliers.}
Note also that because the flux, redshift and stellar-mass
distributions are non-gaussian, it is not wise to use their medians to
calculate the derived quantities \updated{`}directly\updated{'}
\updated{(e.g.~<SSFR>=<SFR>/<$M_*$>)}; instead the full distributions
should be used and the medians calculated as the final step. Our
adopted error on the median (`median absolute deviation', or MAD), is
related to the standard error on the mean $\sigma$ by
$\sigma_{<x>}=\sigma_{\bar{x}}/1.4826$ \citep{hoaglin1983}. We note,
however, that the central-limit theorem does not strictly
apply\updated{.}

During the extraction of fluxes from the radio map, we also excluded
regions of the radio map for which the local rms noise is greater than
$40\,\mu$Jy, the noise map having been generated following the method
of \cite{bondi2003}. In order to test for systematic effects that may
have led to non-zero flux in the single-pixel stacks, we repeated our
stacking analysis as before, but randomizing the positions of 40,000
sources; that process was then itself repeated to give a total of 300
realisations. The resulting measurement of the median simulated flux
was $-0.010\pm 0.007\,\mu$Jy.

\begin{figure*}
\centering
\subfigure[]{
\includegraphics[width=8.5cm,origin=br,angle=0]{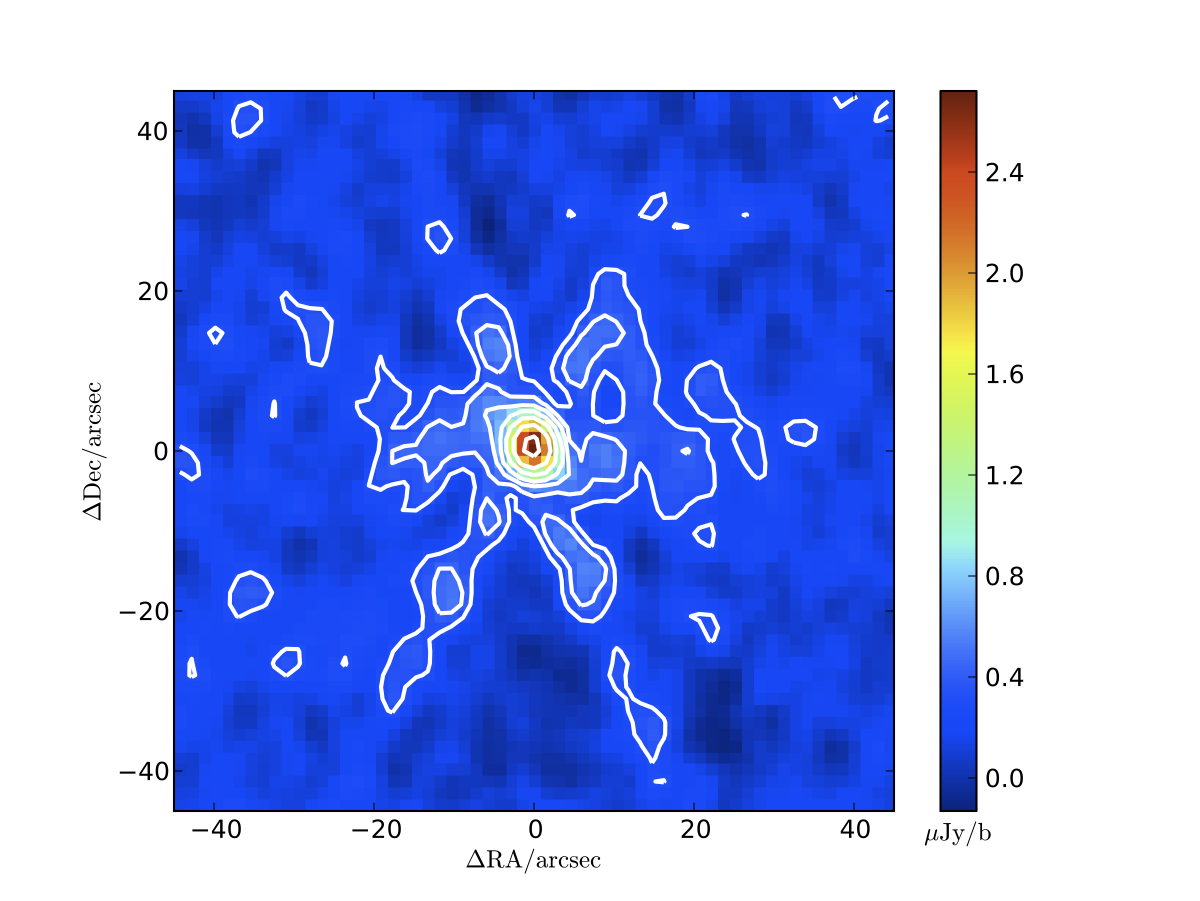}\label{fig:postage:mean}}
\subfigure[]{
\includegraphics[width=8.5cm,origin=br,angle=0]{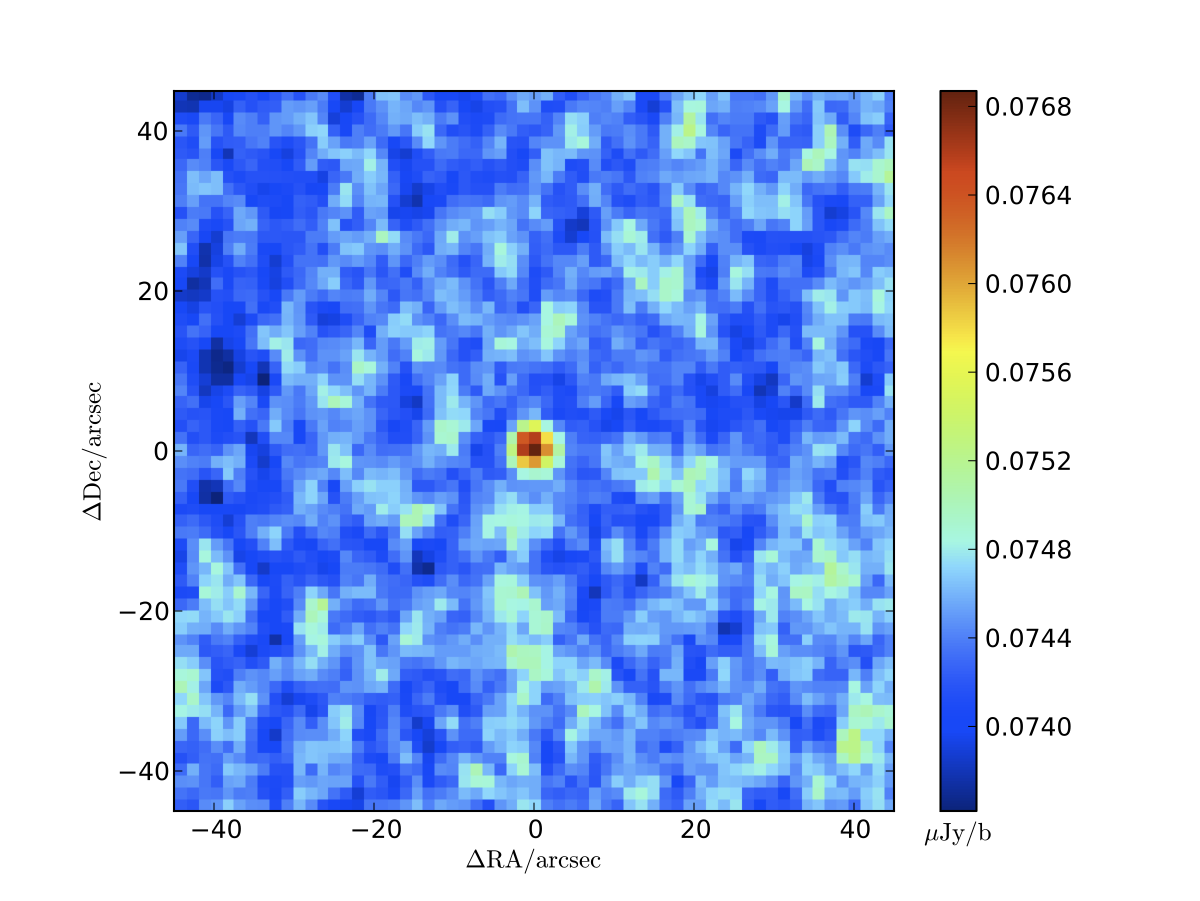}\label{fig:postage:mean_std}}\\
\subfigure[]{
\includegraphics[width=8.5cm,origin=br,angle=0]{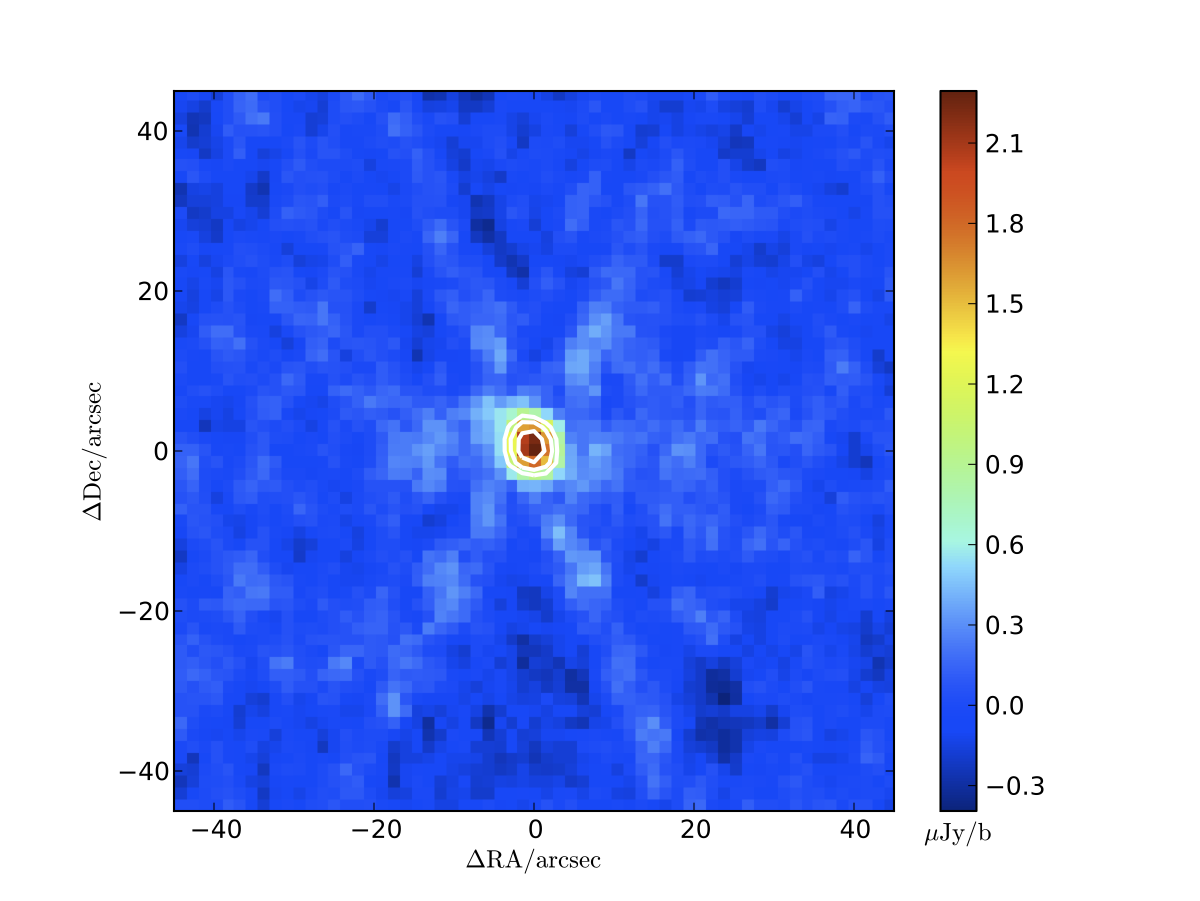}\label{fig:postage:median}}
\subfigure[]{
\includegraphics[width=8.5cm,origin=br,angle=0]{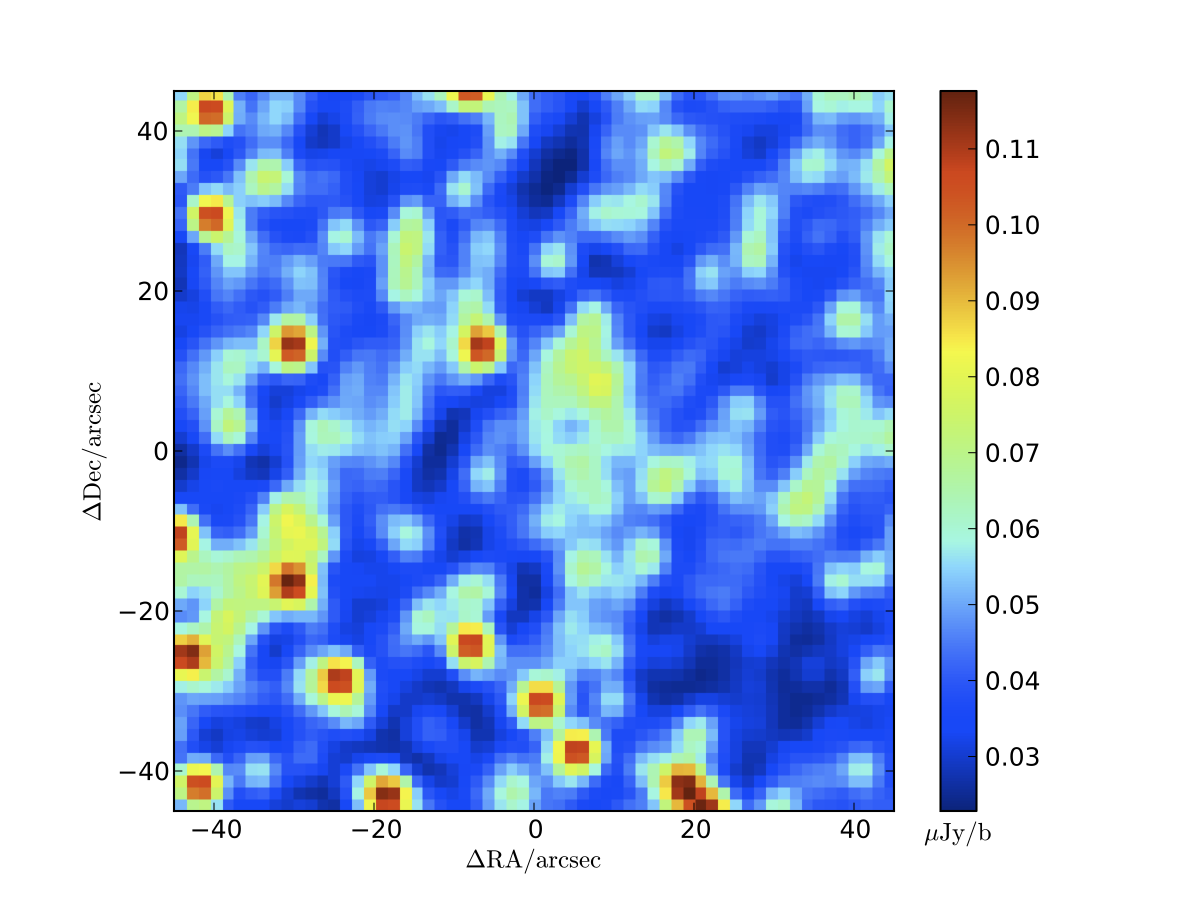}\label{fig:postage:median_std}}
\caption{(a) Noise-weighted stacked 1.4-GHz image and (b) its
  standard-error map. (c) median stacked image and (d) its
  standard-error map. The un\textsc{clean}ed synthesized beam is
  evident in (a) and (c). (b) shows how the map noise will be
  conservative if measured at the central pixel. We have confirmed
  that the noise structure of (d) is consistent with gaussian
  noise. Contour levels begin at $\pm 4\sigma$ and increase by a
  factor of $\sqrt{2}$ thereafter. The $41\times 41$-pixel images have
  a scale of 1.5 arcsec/pixel.\label{fig:postage}}
\end{figure*}

As well as stacking individual pixel fluxes, we have also co-added
$41\times41$ pixel, average flux maps aligned on the positions of the
\Ks-selected sources in each sample, in order to provide visual
confirmation of our procedure. We compute the weighted-mean
\textit{and} median maps and their corresponding standard-error
maps. The synthesized beam, below the threshold for \textsc{clean}ing
the 1.4-GHz map, is now evident in both the weighted-mean (Figure
\ref{fig:postage:mean}) and median (Figure \ref{fig:postage:median})
maps. Figures \ref{fig:postage:mean_std} and
\ref{fig:postage:median_std} respectively show the standard-error maps
corresponding to the values in the weighted-mean and median maps.
Figure \ref{fig:postage:mean_std} implies that any uncertainty
estimated from a central (i.e.~single) pixel will be conservative
compared to a method in which the noise is measured, say, far from the
centre of the map. We have found that the distribution of pixels in
Figure \ref{fig:postage:median_std} is consistent with gaussian noise.

\section{Results}
\label{sec:results}

\subsection{Calculating Specific Star-Formation Rates as a Function of
  Galaxy Type, Redshift and Stellar Mass}
\label{sec:sfrs}

Following \citet{dunne2009}, for each pixel in the radio map where we
have a flux $S_{1.4}$ we determine\referee{d} the rest-frame 1.4-GHz
luminosity $L_{1.4}$ assuming a spectral index $\alpha=-0.8$. We
calculate\referee{d} the SFR by following \cite{condon1992},
\cite{haarsma2000}, \cite{condon2002} and \cite{dunne2009}\updated{:}

\begin{equation}
\label{eqn:conversion:condon}
\left(\frac{\mathrm{\referee{SFR}}}{M_{\odot}\mathrm{yr}^{-1}}\right) = 1.2006 \times 10^{-21} \,\left(\frac{L_{1.4}}{\mathrm{WHz}^{-1}}\right).
\end{equation}

\noindent The SSFR for each object is then the SFR divided by the
stellar mass.

\begin{table}
\centering
\caption{Summary of object samples \updated{($0<z<3$)}. The columns are: (1) Galaxy type; (2)
  total number of objects in each stack (including 5-$\sigma$
  detections); (3) fraction of sources detected at
  \updated{$>5\sigma$} in the radio data; and (4) median redshift.\label{table:species}}
\begin{tabular}{lccc}
\hline
Sample \T & Number & Detected/\% & $<z>$ \\
(1) \B & (2) & (3) & (4) \\
\hline
all \T\B & \updated{49604} & 0.55 & \updated{1.348} \\
elliptical & \updated{9900} & 1.47 & \updated{1.442} \\
irregular & \updated{33747} & 0.39 & \updated{1.334} \\
starburst & \updated{5957} & 0.20 & \updated{1.207} \\
star & 114 & 0.89 & -- \\
millennium \B & 7388770 & -- & 1.250 \\
\hline
\end{tabular}
\end{table}

In this \updated{analysis} we report our results for the dependence of
SSFR on stellar mass, as well as for its evolution with redshift.

\subsubsection{Separation of SSFR Dependence}
\label{sec:results:ssfrs:sepn}

\updated{In order to} quantify the relationship between the SSFR and
each of $M_*$ and $z$, we first note (as do \citealt{karim2011}) the
assumption suggested by the data that the functional dependence of
SSFR on each quantity is separable (uncorrelated), i.e.~

\begin{equation}
\label{eqn:ssfr_separate}
\mathrm{\referee{SSFR}}\left(M_*,z\right) \propto
\mathrm{SSFR}\left(\left. M_*\right|_z\right)  \mathrm{SSFR}\left(\left.z\right|_{M_*}\right) = M_*^{\beta} (1+z)^n.
\end{equation}

\noindent We therefore fit\referee{ted} these two separate functions
of $M_*$ and $z$ using a weighted least-squares estimator:

\begin{equation}
\label{eqn:ssfr_M_given_z}
\mathrm{\referee{SSFR}}\left(\left. M_*\right|_z\right) = c\left(z\right)  M_*^{\beta},
\end{equation}

\noindent

\begin{equation}
\label{eqn:ssfr_z_given_M}
\mathrm{\referee{SSFR}}\left(\left.z\right|_{M_*}\right) = C\left(M_*\right)  (1+z)^n.
\end{equation}

\noindent In what follows \updated{we} examine the relationship
between SSFR, $M_*$ and $z$. \referee{Note that a typical $M_*$--$z$
  bin contains at least 100 objects.}

\subsubsection{Dependence on Stellar Mass}
\label{sec:results:ssfrs:M}

\begin{figure*}
\subfigure{
\includegraphics[width=8cm,trim=20mm 0mm 20mm 0mm,origin=br,angle=0]{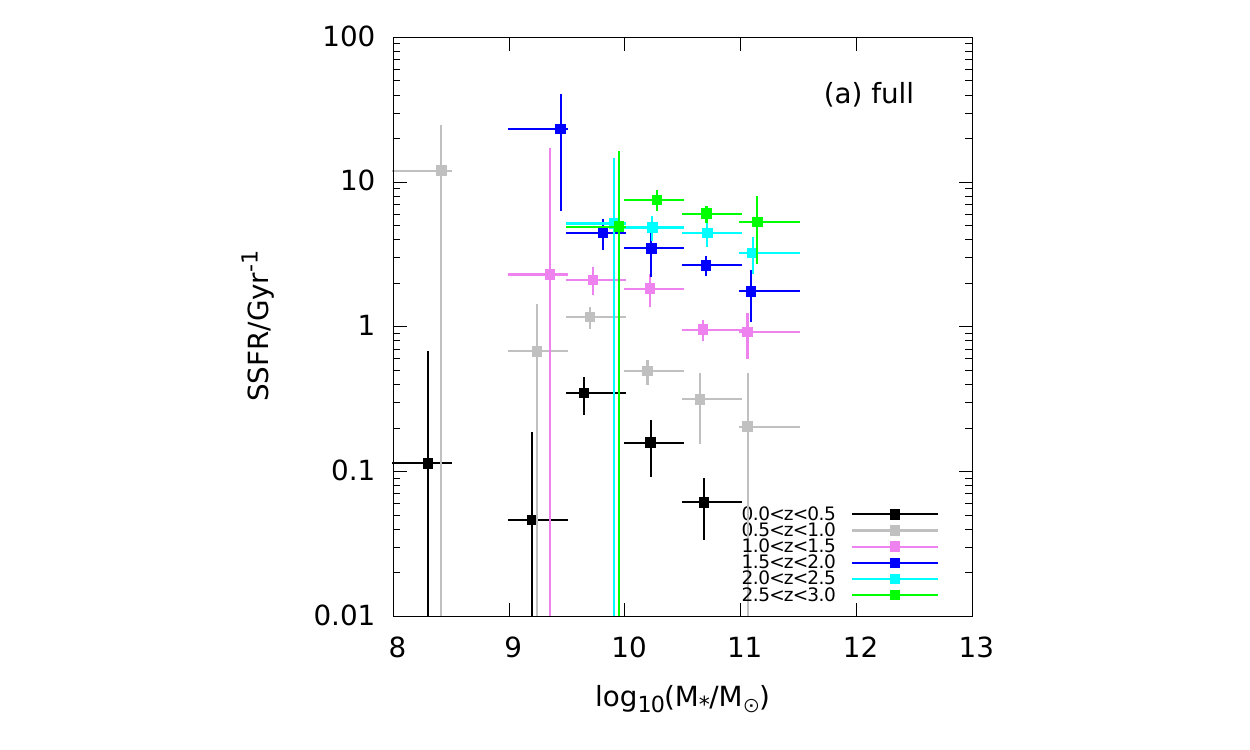}\label{fig:ssfrs:all_v_M}}
\subfigure{
\includegraphics[width=8cm,trim=20mm 0mm 20mm 0mm,origin=br,angle=0]{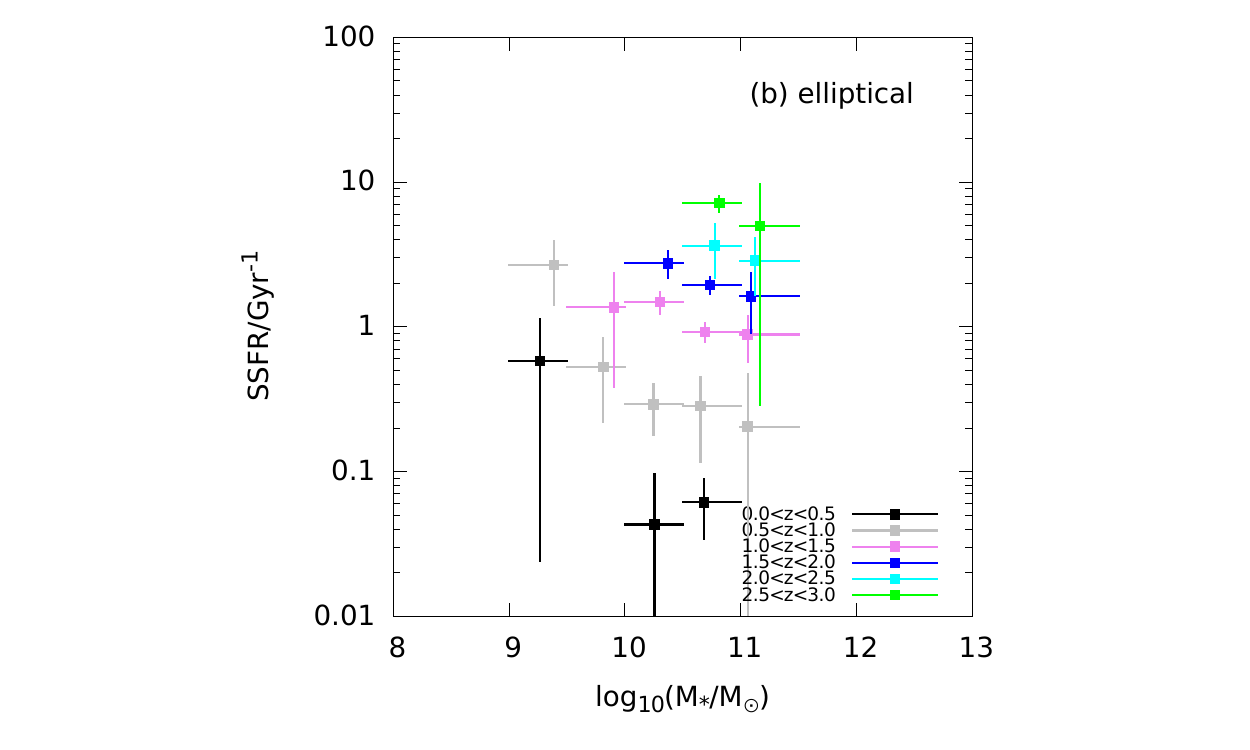}\label{fig:ssfrs:ell_v_M}}
\subfigure{
\includegraphics[width=8cm,trim=20mm 0mm 20mm 0mm,origin=br,angle=0]{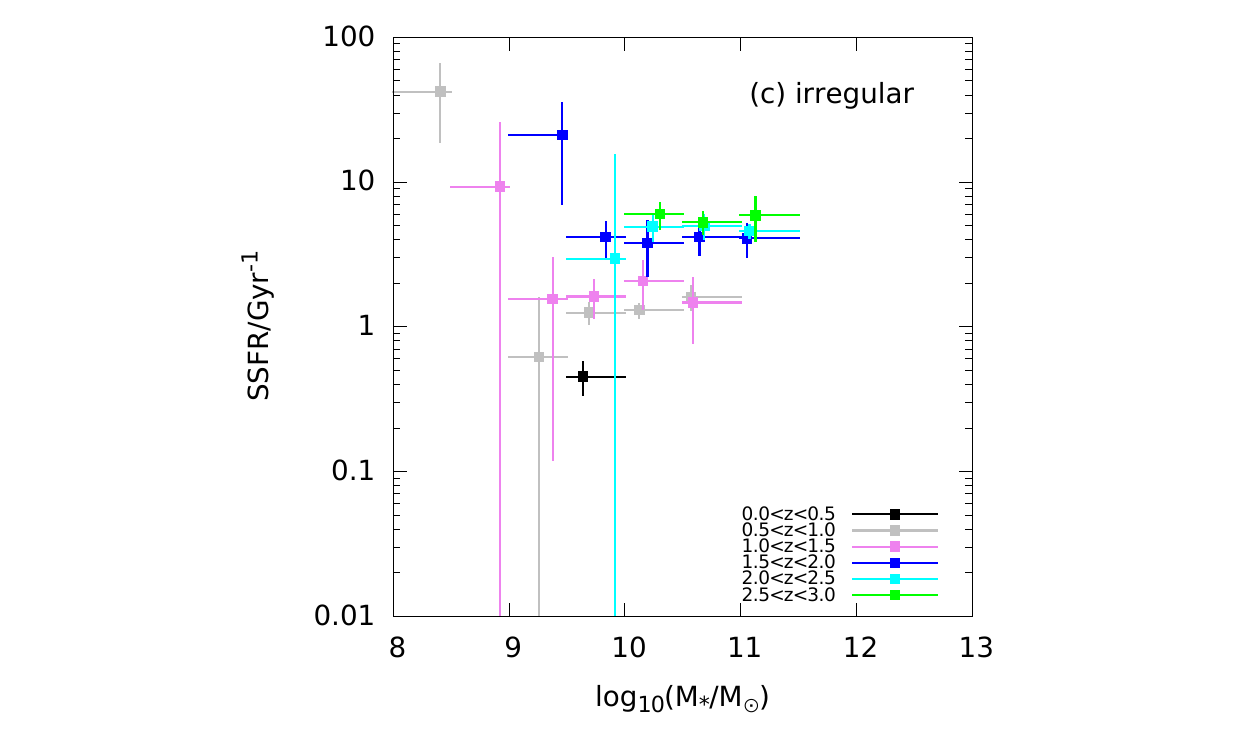}\label{fig:ssfrs:mid_v_M}}
\subfigure{
\includegraphics[width=8cm,trim=20mm 0mm 20mm 0mm,origin=br,angle=0]{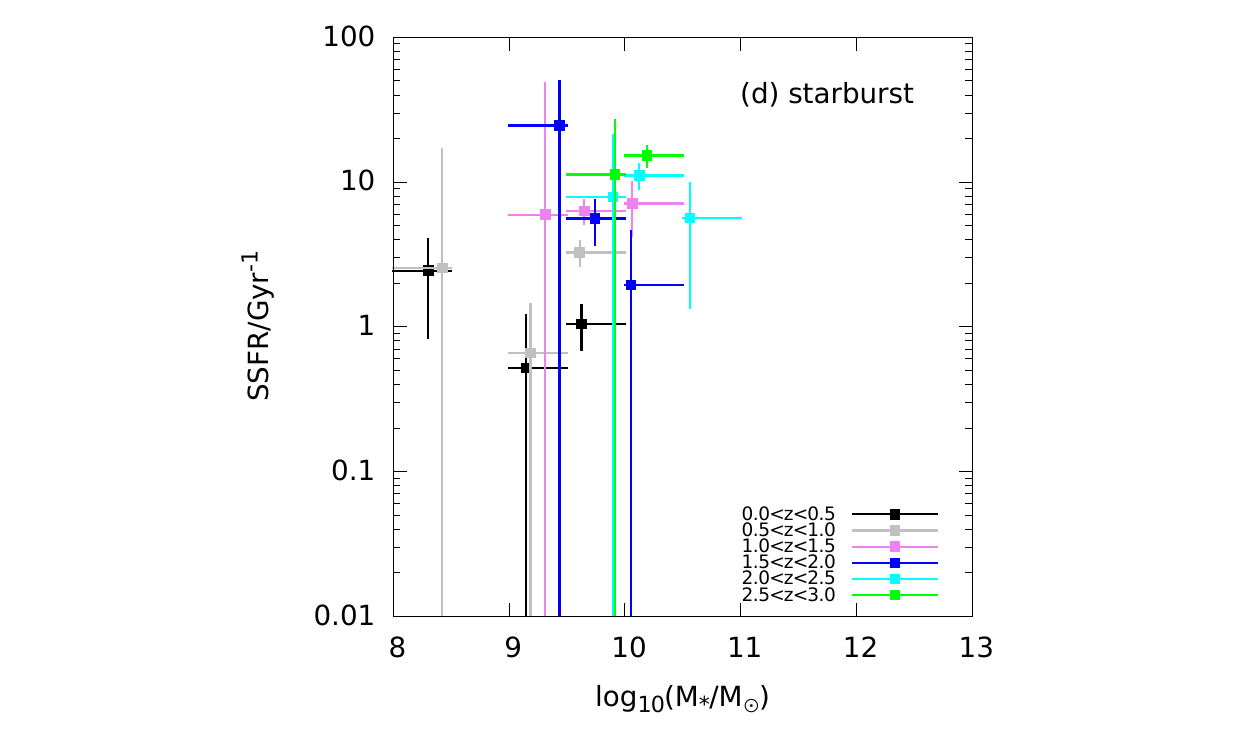}\label{fig:ssfrs:sbn_v_M}}
\caption{Specific star-formation rate as a function of stellar mass at
  a given redshift for (a) all galaxies, (b) elliptical galaxies, (c)
  irregular galaxies and (d) starbursts. SSFR tends to decrease as
  stellar mass increases, independently of redshift.\label{fig:ssfrs-M}}
\end{figure*}

In Figures~\ref{fig:ssfrs:all_v_M}--\ref{fig:ssfrs:sbn_v_M} we show
the dependence of SSFR on stellar mass for the different samples: all
galaxies, ellipticals, irregulars and starbursts. As a basic point, we
note that SSFRs are significantly higher for the starburst sample than
for the other samples. One can also see straightaway the general trend
that SSFR decreases with increasing stellar mass.

\begin{figure*}
\centering
\subfigure{
\includegraphics[trim = 20mm 0mm 20mm 0mm, width=8.5cm,origin=br,angle=0]{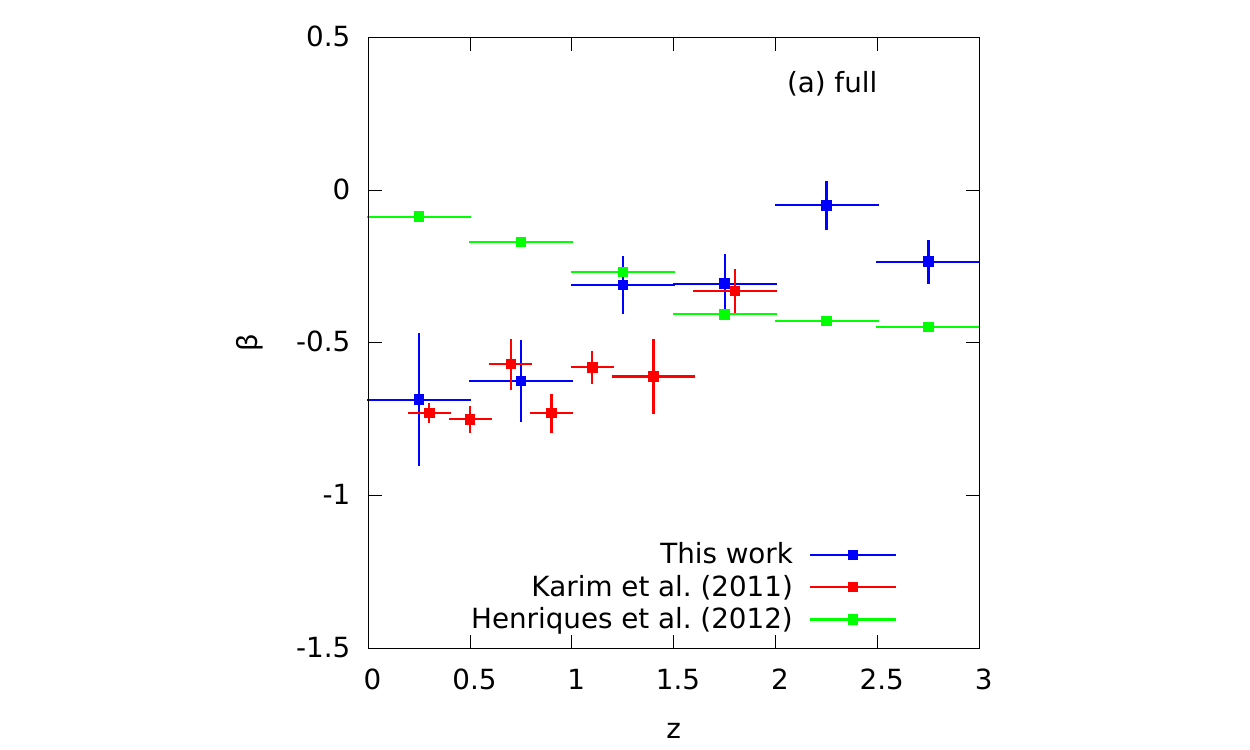}\label{fig:beta-compare:all}}
\subfigure{
\includegraphics[trim = 20mm 0mm 20mm 0mm, width=8.5cm,origin=br,angle=0]{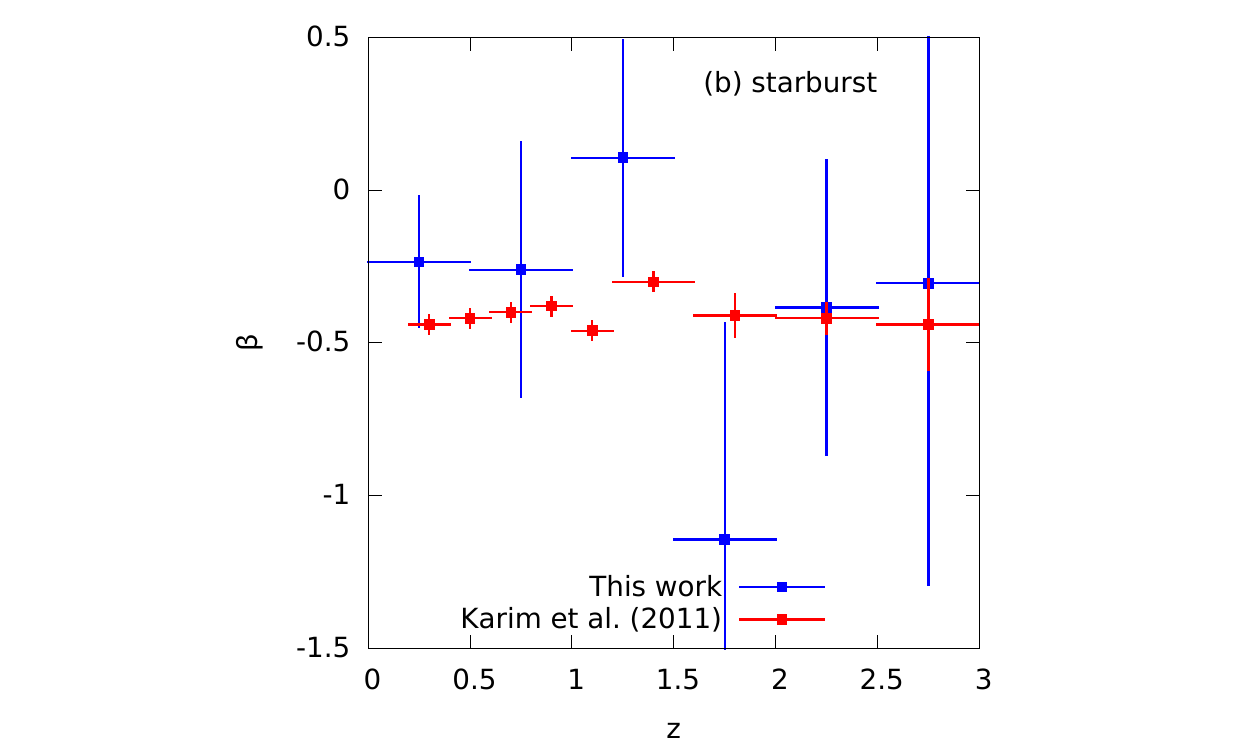}\label{fig:beta-compare:sbn}}
\caption{Comparison of \updated{mass gradient} $\beta$ of SSFR against
  stellar mass
  as a function of redshift for the (a) full and (b) starburst
  samples\updated{.}
  \label{fig:beta-compare}}
\end{figure*}

For the full sample, the `mass gradient' $\beta_{all}$ is negative in
all cases, but flattens off with redshift
(Figure~\ref{fig:beta-compare:all}), i.e.~the steepness of SSFR with
stellar mass is lower at higher redshift
($\mathrm{d}\beta/\mathrm{d}z>0$). We return to this point in section
\ref{sec:discussion}. In contrast, for the starburst sample, SSFR is
relatively constant with stellar mass
(Figure~\ref{fig:beta-compare:sbn}). \updated{The mass} gradient is
also shallower for the starburst sample compared to all galaxies,
especially at $z\lesssim 1$.

\begin{figure}
\centering
\includegraphics[trim = 20mm 0mm 20mm 0mm, width=8.5cm,origin=br,angle=0]{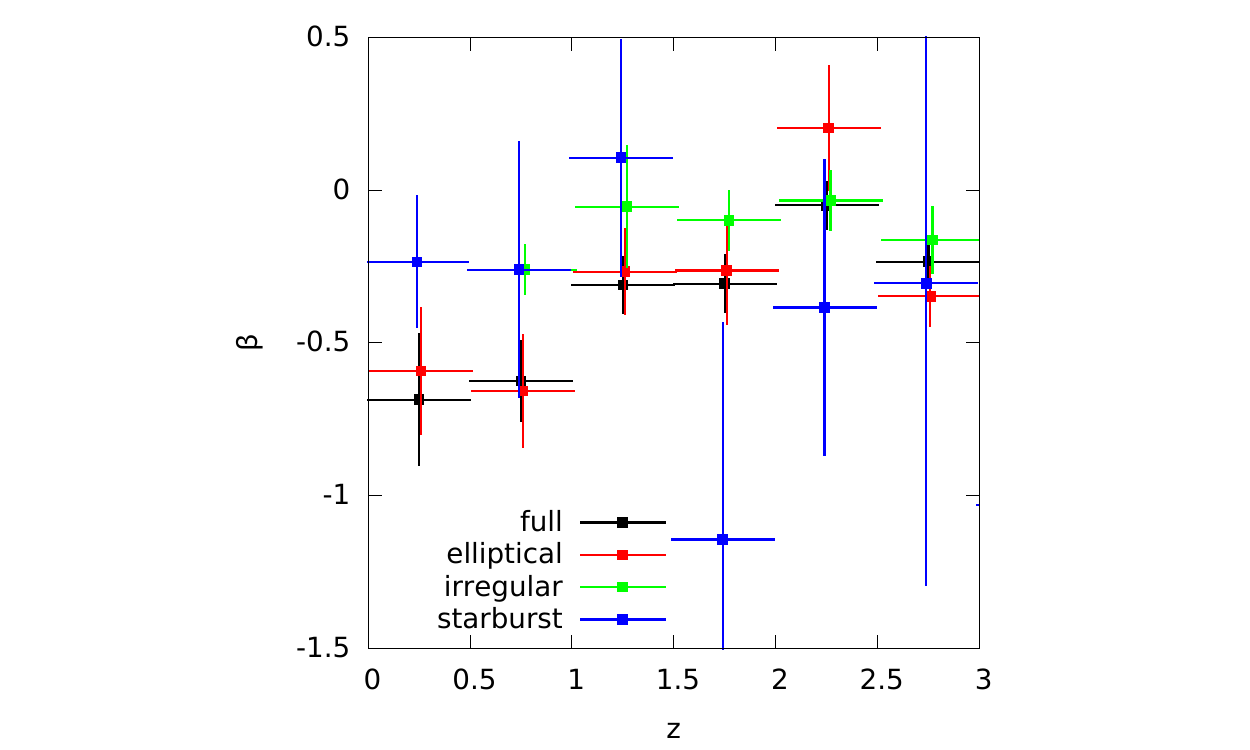}
\caption{Comparison of gradients $\beta$ of SSFR as a function of
  stellar mass at given redshifts for all of our samples\updated{.}
  \label{fig:beta-compare:every}}
\end{figure}

The trend for ellipticals (Figure~\ref{fig:beta-compare:every}) is
almost the same as that for the full sample, confirming that the
former population is the dominant contributor to the latter. Just as
for $\beta_{sbn}$, $\beta_{irr}$ is less negative than $\beta_{ell}$
and $\beta_{all}$, and roughly independent of redshift, implying that
irregular and starburst galaxies are at the same stage of evolution at
all redshifts probed. However, for ellipticals, there is evolution,
with the low-redshift ellipticals having much lower SSFRs at the
high-mass end. In other words, star formation has terminated earlier
for ellipticals.

\subsubsection{Dependence on Redshift}
\label{sec:results:ssfrs:z}

\begin{figure*}
\centering
\subfigure{
\includegraphics[width=8.5cm,trim=20mm 0mm 20mm 0mm,origin=br,angle=0]{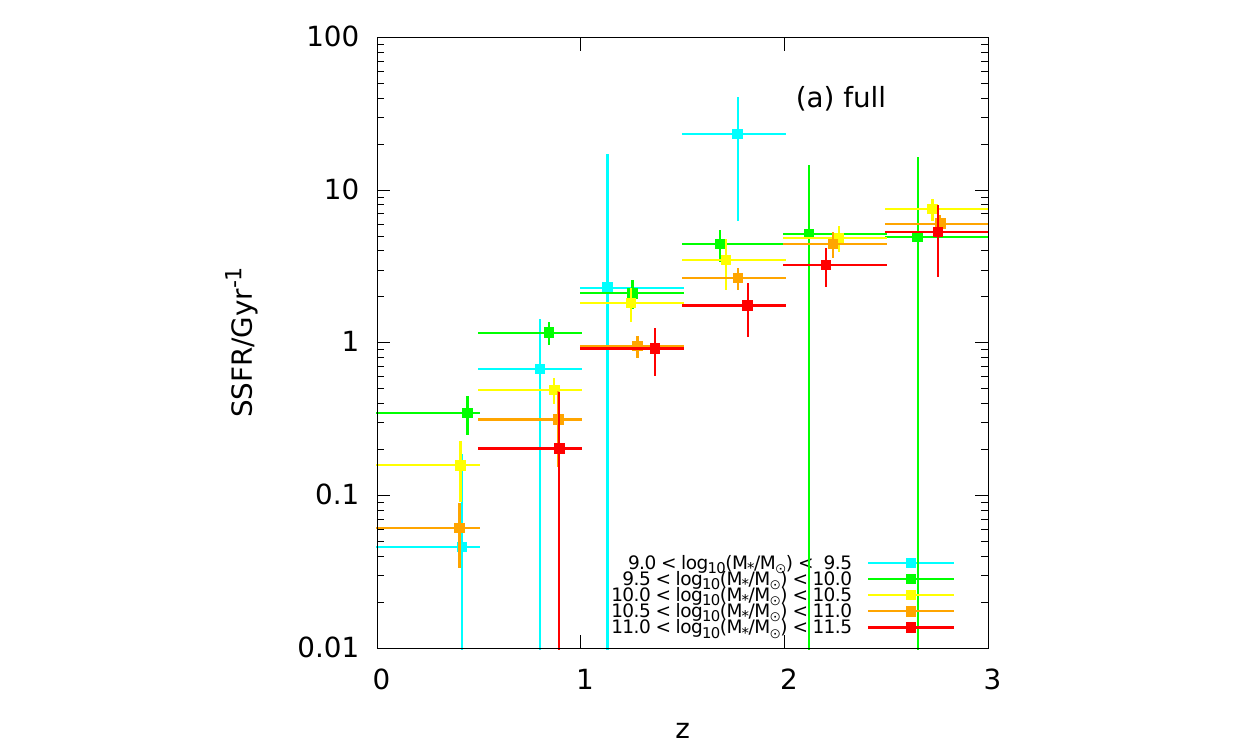}\label{fig:ssfrs:all_v_z}}
\subfigure{
\includegraphics[width=8.5cm,trim=20mm 0mm 20mm 0mm,origin=br,angle=0]{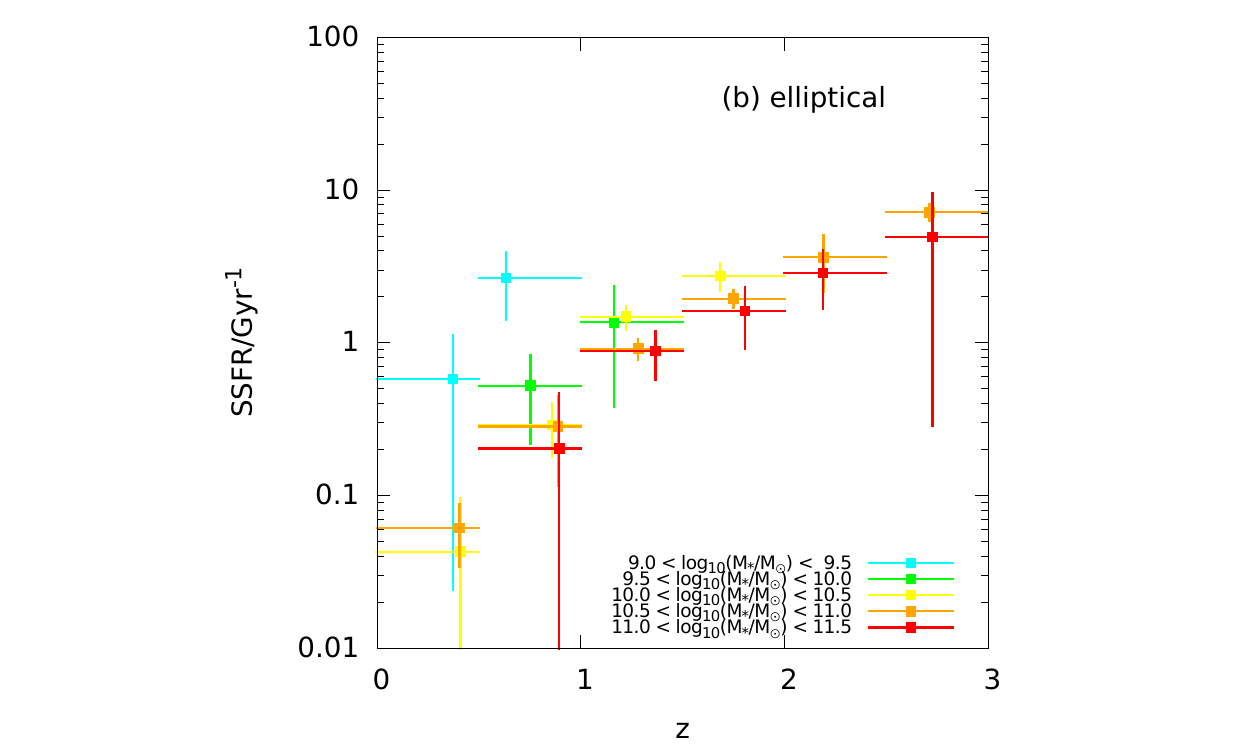}\label{fig:ssfrs:ell_v_z}}
\subfigure{
\includegraphics[width=8.5cm,trim=20mm 0mm 20mm 0mm,origin=br,angle=0]{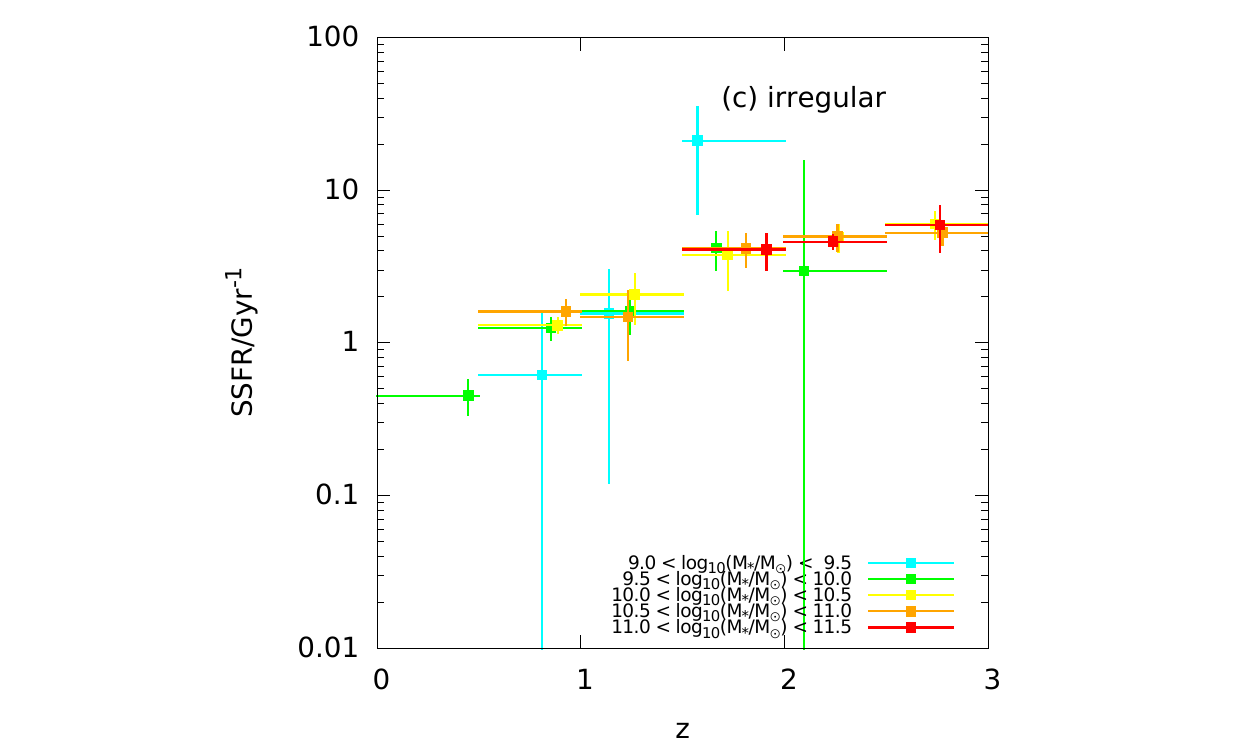}\label{fig:ssfrs:mid_v_z}}
\subfigure{
\includegraphics[width=8.5cm,trim=20mm 0mm 20mm 0mm,origin=br,angle=0]{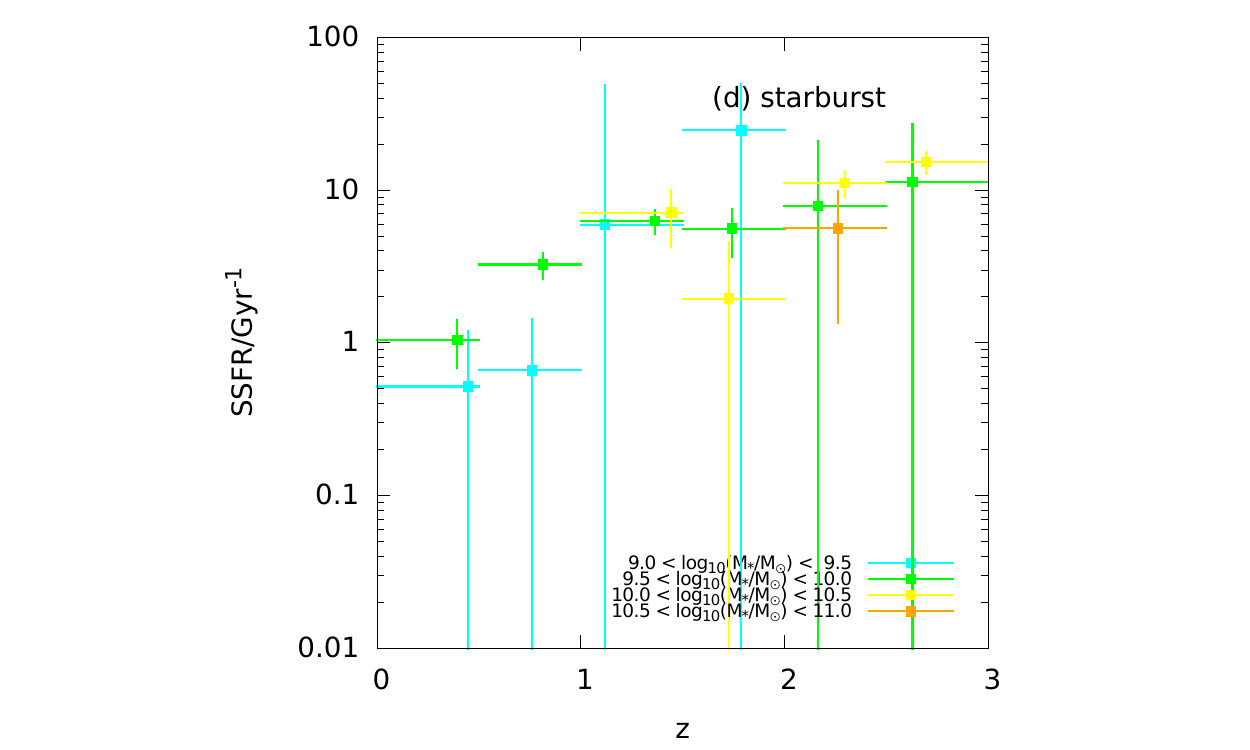}\label{fig:ssfrs:sbn_v_z}}
\caption{Specific star-formation rate as a function of redshift at a
  given stellar mass for (a) all galaxies, (b) ellipticals, (c)
  irregular galaxies and (d) starburst galaxies. SSFR increases with
  redshift.\label{fig:ssfrs-z}}
\end{figure*}

Figures~\ref{fig:ssfrs:all_v_z}--\ref{fig:ssfrs:sbn_v_z} indicate how
SSFRs for our samples evolve with redshift. One possible definition
(used by e.g.~\citealt{elbaz2011}) of starburst galaxies is those
objects for which the SSFR is $>2$ times the typical `main sequence'
SSFR. This definition is consistent with our results (Figures
\ref{fig:ssfrs:all_v_z} and \ref{fig:ssfrs:sbn_v_z}) that galaxies
classified as starbursts from our SED fitting have SSFRs that are
greater than approximately a factor of two higher than SSFRs for
galaxies in the full sample (see also section
\ref{sec:results:comparison:elbaz11}).

Figures \ref{fig:ssfrs:mid_v_z}/\ref{fig:ssfrs:sbn_v_z} and Figures
\ref{fig:ssfrs:all_v_z}/\ref{fig:ssfrs:ell_v_z} show the redshift
evolution for irregulars and the starburst sample, and for the full
sample and the ellipticals, respectively. Compared to the full sample,
the \updated{gradients for the redshift evolution of the starbursts}
are shallower (1--5/Gyr$^{-1}$ versus 0.1--5/Gyr$^{-1}$), i.e.~0.5
decades to 1.4 decades, up to $z\approx 2.5$. This implies that the
evolution of the SSFR over cosmic time is much faster for ellipticals
than for starburst galaxies, i.e.~SSFR in the starburst sample holds
up at late times compared to the elliptical sample. We see therefore
that more massive galaxies formed their stars earlier than less
massive ones, which is consistent with the `downsizing' scenario
\citep{cowie1996,pg2008}. SSFR at a given stellar mass then tends to
flatten out by and above $z\approx 2.5$, although our data are limited
at higher redshift.

\begin{figure}
\centering
\subfigure{
\includegraphics[trim = 20mm 0mm 20mm 0mm, width=8.5cm,origin=br,angle=0]{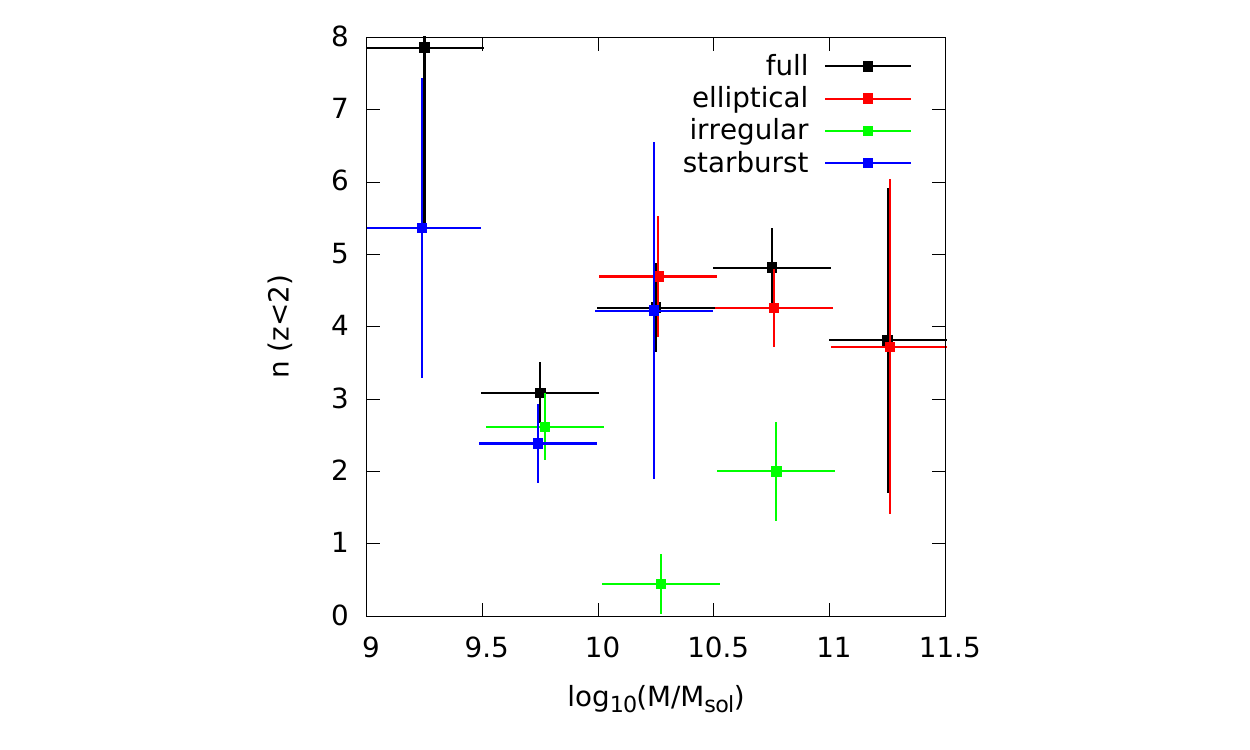}\label{fig:n-compare:nz2}}
\subfigure{
\includegraphics[trim = 20mm 0mm 20mm 0mm, width=8.5cm,origin=br,angle=0]{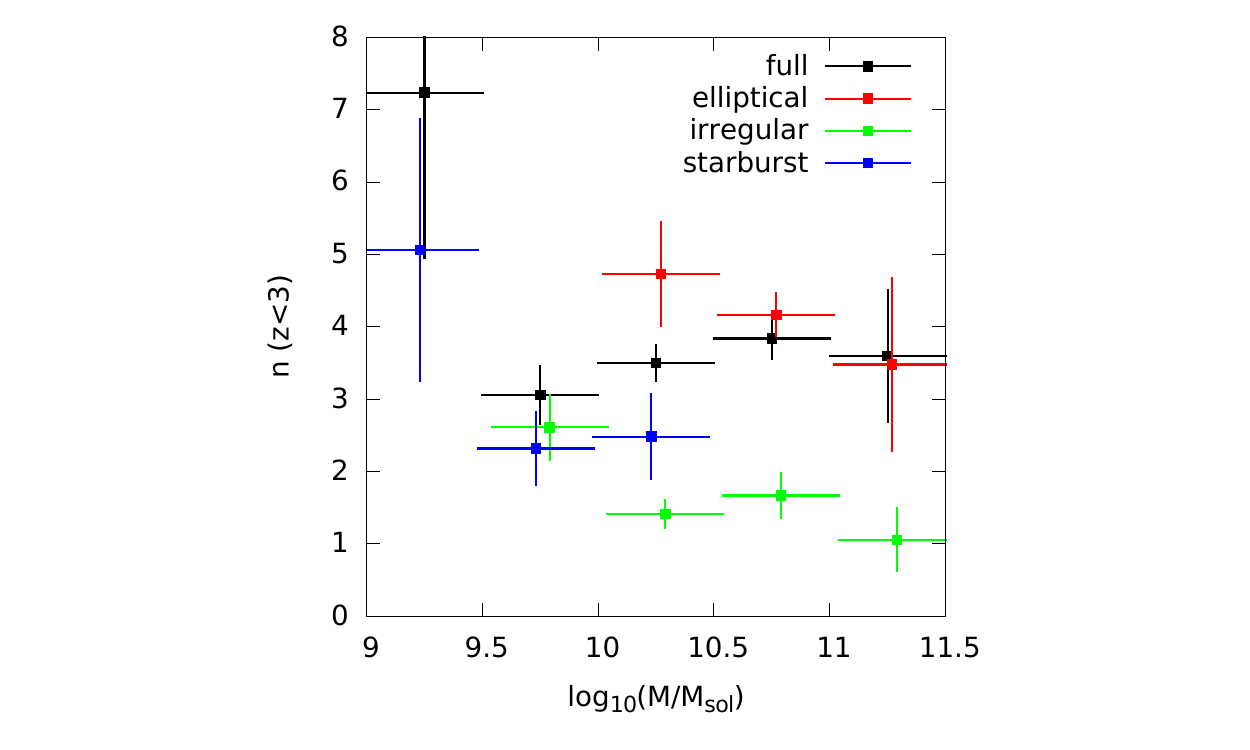}\label{fig:n-compare:nz3}}
\caption{Comparison of redshift-evolution parameters $n$ of SSFR as a
  function of stellar mass for all of our samples. \updated{\textbf{Upper panel:}
  $0<z<2$ and \textbf{Lower panel:} $0<z<3$\updated{.}}
\label{fig:n-compare-quick}}
\end{figure}

The redshift-evolution parameter $n$ (see
Figure~\ref{fig:n-compare-quick}) is on average higher for all
galaxies than for the starburst sample, meaning that, at a given
stellar mass, redshift evolution is stronger for the full sample than
for the starburst sample. Figure~\ref{fig:n-compare-quick} also shows
that the redshift evolution $n$ is very distinct for the three
different subsamples: the $n$ values for the starburst and irregular
samples are consistent with being independent of mass, but the
ellipticals exhibit a falling gradient with increasing $M_*$
($\mathrm{d}n/\mathrm{d}M<0$). Hence star formation in the
highest-mass ellipticals does not evolve as much as for the
lowest-mass ellipticals\updated{, implying that the morphology as well
  as mass plays a role in the evolution of the star-formation rate density.}

\section{Comparison with SSFRs from Other Routes}
\label{sec:discussion}

In order to set our results in context, we now compare them to those
from other work. We have divided the comparison into three
subsections: (i) those where \Ks-band sources have been stacked at
1.4\,GHz, (ii) simulations, and (iii) measurements made using
far-infrared SFR indicators. Note that in several cases authors have
considered more than one SFR indicator, so a clean separation is not
always possible.

\subsection{SSFRs from Radio Stacking}
\label{sec:results:comparison:karim11}

\subsubsection{Dependence on Stellar Mass}
\label{sec:results:comparison:karim11:mstar}

We have found that SSFR decreases with stellar mass (downsizing;
Figure \ref{fig:ssfrs-M}) and increases with redshift (Figure
\ref{fig:ssfrs-z}), and that the gradients $\beta$, though all
negative, become shallower with increasing redshift
($\mathrm{d}\beta/\mathrm{d}z>0$). Our results therefore agree
qualitatively with those of \cite{karim2011}, out to $z\approx
2$. Comparing \cite{karim2011}'s Figure 6 with our Figures
\ref{fig:ssfrs-M} and \ref{fig:ssfrs-z}, our SSFRs are slightly higher
than theirs for the starbust sample, but the dynamic ranges are of the
same order for both the starburst sample and for all galaxies.

Quantifying, Figure \ref{fig:beta-compare:all} includes a comparison
of our measurements of $\beta$ as a function of redshift for the full
sample. Extending the study of \cite{karim2011} to higher redshift, we
find that $\beta$ continues to flatten with redshift with
approximately the same gradient $\mathrm{d}\beta/\mathrm{d}z$
($\gtrsim 0$).

Figure \ref{fig:beta-compare:sbn} shows $\beta\left(z\right)$ for the
starburst sample; the mass slope $\beta_{sbn}$ is independent of
redshift, with a weighted-mean slope $\beta_{\mathrm{w}}$ of
$-0.44\pm0.18$), in agreement with the \cite{karim2011}
findings. (Their Table 4 refers; it is not clear whether their
`standard error' is that on the mean or not, but our conclusion is
unaffected.) \cite{karim2011} concluded for a starburst sample that
$\beta$ is independent of redshift, but note, as do \cite{elbaz2011},
that the level and shape of $\beta(z),$ including any potential
`upsizing' (\citealt{rodighiero2010}, \citealt{oliver2010}), is a
strong function of how such starburst objects are selected: negative
for a mildly star-forming population but `flat'
(i.e.~$\mathrm{d}\beta/\mathrm{d}z=0$) for actively star-forming
galaxies. \cite{karim2011} only use 21 SEDs (to our 62) in determining
redshifts and stellar masses, with an additional colour-colour cut. We
therefore advocate that our results for the mass slopes are consistent
with those of \cite{karim2011}, but our data do not permit us to
investigate the upsizing scenario (i.e.~that $\beta$ steepens with $z$
at $z\gtrsim 3$) seen by \cite{rodighiero2010} and \cite{oliver2010}.

The SSFRs measured by \cite{dunne2009}, derived using the Condon
calibration, are higher than ours for the star-forming galaxies, which
emphasizes the different selection functions and methodologies. For
example, we use the complete SEDs to estimate stellar masses, whereas
\cite{dunne2009} use only the $K$-band absolute magnitude. They find
that SSFR increases strongly with redshift, for all galaxies and all
stellar-mass bins, except that the SSFRs in the higher-mass bins
flatten off at early times. This `bimodality' in the SSFR-$M_*$ plane,
with excess `red and dead' galaxies at higher stellar masses (in other
words, extrapolation to lower stellar masses generally overpredicting
SSFRs), is also seen by \cite{karim2011}, but is only hinted at in our
data in the lowest two redshift bins. \cite{dunne2009} further found
that SSFR decreases slowly with stellar mass, and that the
stellar-mass slope $\beta_{all}$ \textit{steepens} with $z$
($\mathrm{d}\beta/\mathrm{d}z<0$); this is in contrast to our findings
and those of \cite{karim2011}. As noted above and in
section~\ref{sec:intro}, this is almost certainly due to the way in
which the \cite{dunne2009} stellar masses have been estimated.

\cite{dunne2009} find that SSFR as a function of cosmic time for their
sBzK sample evolves less than for a non-BzK sample. We find the same
if we equate our starburst sample with their sBzK, and our full sample
with their non-BzK sample. The fact that the stellar-mass estimation
by \cite{dunne2009} is via the rest-frame K-band absolute magnitude
(cf.~\citealt{serjeant2008}), rather than by the SED-fitting method
shared by our work and that of \cite{karim2011}, is the sole major
methodical difference that might explain the apparent
discrepancy. \cite{dunne2009} note the potential introduction of
systematics by their SFR--$M_*$ conversion method and warn against
overinterpretation.

Rather than using SED fitting, \cite{pannella2009} selected
star-forming sBzK galaxies at $z\approx2$, deriving SFRs from stacked
VLA 1.4-GHz data over the COSMOS field and paying particular attention
to cleaning AGN from their sample. \cite{pannella2009} find the slope
$\beta_{\mathrm{sBzK}}$ is zero (flat) at $z\approx2$ from 1.4-GHz
data, but negative in the UV, i.e.~more massive galaxies have SEDs
that may be subject to greater extinction. This highlights the problem
of dust attenuation for UV star-formation tracers, and so, as
expected, our results at $z\approx2$ are consistent with their
radio-stacking results but inconsistent with their UV index, although
it may be possible to reconcile the two by including the effects of
extinction.

\begin{figure}
\centering
\includegraphics[trim = 20mm 0mm 20mm 0mm,width=8.5cm,origin=br,angle=0]{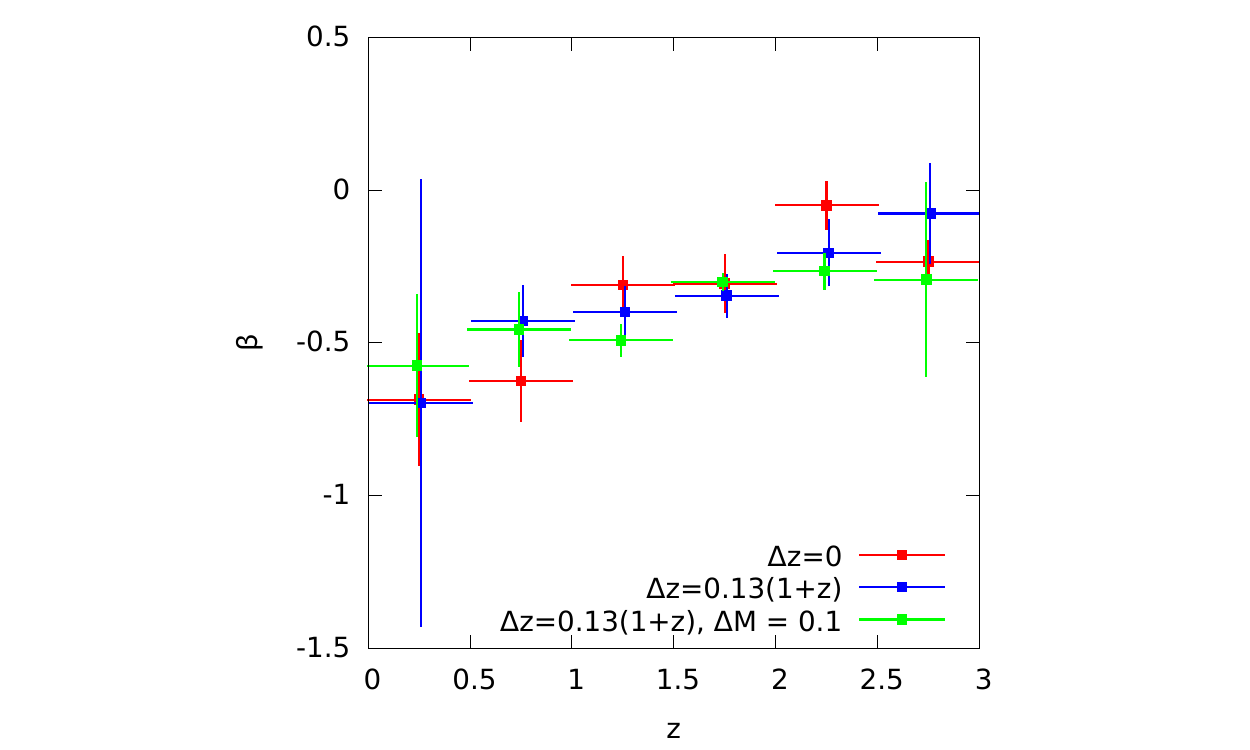}
\caption{\referee{Mass gradient $\beta$ as a function of redshift for
    the full sample with (blue) and without (red) the effect of
    photometric-redshift errors included. The green points further
    include a contribution from stellar-mass uncertainties. There is
    no discernable difference in either the values nor in the shape of
    $\beta(z)$.}\label{fig:scrambled-zphots}}
\end{figure}

\referee{In section \ref{sec:seds:zphots} we asserted that our results
  would not be biased by neglecting the typical photometric-redshift
  uncertainties of $\Delta z/(1+z)$=0.13.
  In order to test this statement, we re-analyzed the
  full-sample data with gaussian uncertainties (i.e.~$\Delta
  z$=$0.13(1+z)$) introduced to the original photometric-redshift
  estimates. We also re-analyzed those data with gaussian
  uncertainties additionally introduced to the original stellar-mass
  estimates ($\log_{10}\left(\Delta M_*/M_{\odot}\right)=0.1$). We
  found no discernable differences in either the values of $\beta$ nor
  in the shape of $\beta(z)$ (Figure~\ref{fig:scrambled-zphots}).}

In summary, we find good agreement among the different radio-derived
SSFRs, as a function of stellar mass, from the literature, and between
those studies and our results, except for the noted discrepancy with
the \cite{dunne2009} SSFR--$M_*$ relation as a result of different
stellar-mass estimation routes.

\subsubsection{Dependence on Redshift}

Redshift evolution of SSFR (measured by $n$; see
Figure~\ref{fig:n-compare-quick}) is much faster for the full sample
than for the starburst sample. Evolution of the irregulars is the
slowest of all our samples, and for the ellipticals $n$ is slightly
steeper than for the full sample at all redshifts.

Ignoring the lowest-mass bin with its low-number statistics, the
slopes for the three different populations are very different: those
for the irregular and starburst samples are consistent with being
independent of mass, whereas the ellipticals exhibit a negative slope
($\mathrm{d}n/\mathrm{d}M < 0$). Hence, star formation in the
higher-mass ellipticals does not evolve as much as in the lower-mass
ellipticals. We further find for all samples that as higher-redshift
objects are included, there is some evidence that the mean level of
$n$ decreases.

\cite{karim2011} find an increasing dependence of $n_{all}$ on stellar
mass ($\mathrm{d}n/\mathrm{d}M>0$), though with a relatively shallow
gradient. Also, their $n$ values for the full and starburst samples
are systematically higher than ours, i.e.~evolution is consistently
faster. \cite{karim2011} note that at $z>1$ their redshift-evolution
parameter $n_{all}$ is similar to the radio-luminosity redshift
dependence of $(1+z)^{3.8}$. SSFR is proportional to $L_{1.4}$
(equation \ref{eqn:conversion:condon}), implying that systematic
errors in the median redshift estimator --- and hence in their results
--- are limited. We do see the same in our full sample (supporting the
robustness of our conclusions), but not for the irregular or starburst
samples.

In conclusion, for both the full and starburst samples, we find that
SSFR--redshift evolution is largely consistent between the VIDEO and
COSMOS data sets, though the evolution parameter $n$ is slightly lower
in our data.

\subsection{Specific Star-Formation Rates for the
  Henriques~et~al.~(2012) Semi-Analytic Model}
\label{sec:results:millennium}

In order to relate our results to simulations, we drew a sample of
galaxies from the semi-analytic model of \cite{henriques2012} that had
been overlaid on the Millennium simulation (\citealt{springel2005},
\citealt{guo2011}). The \cite{henriques2012} model was specifically
designed for comparison with deep, high-redshift surveys such as
VIDEO. Each pencil beam has an area of $1.4\times1.4$ square degrees
and includes all the observables required: photometry from $4000\AA$
right through to 6\,$\mu$m, redshifts, star-formation rates and
stellar masses. Note that because of the resolution of the
\cite{henriques2012} model, the usable stellar mass is restricted to
$M_*>10^9M_{\odot}$. Further, just as for the VIDEO data set, we also
restricted the \cite{henriques2012} sample to $\Ks<23.5$.

\begin{figure*}
\centering
\subfigure[]{
\includegraphics[trim = 30mm 0mm 20mm 0mm, height=7.5cm,origin=br,angle=0]{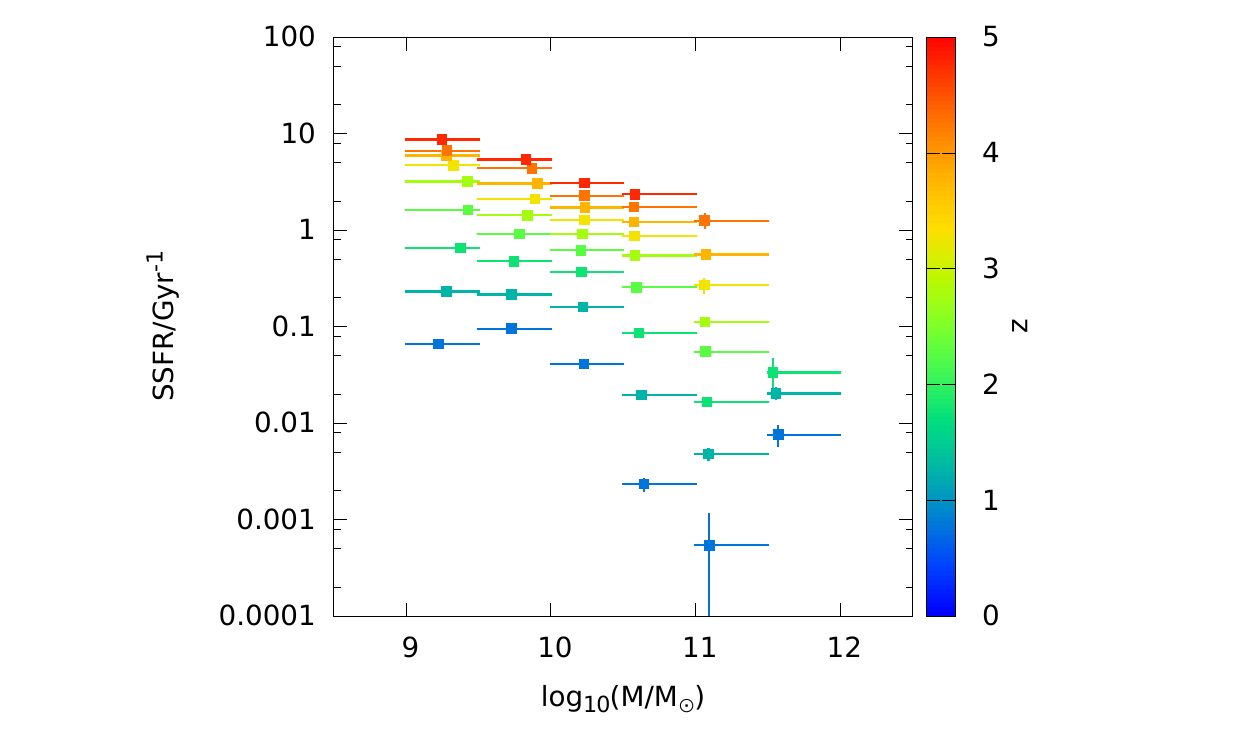}\label{fig:mill:SSFR_v_M}}\quad\quad
\subfigure[]{
\includegraphics[trim = 20mm 0mm 30mm 0mm, height=7.5cm,origin=br,angle=0]{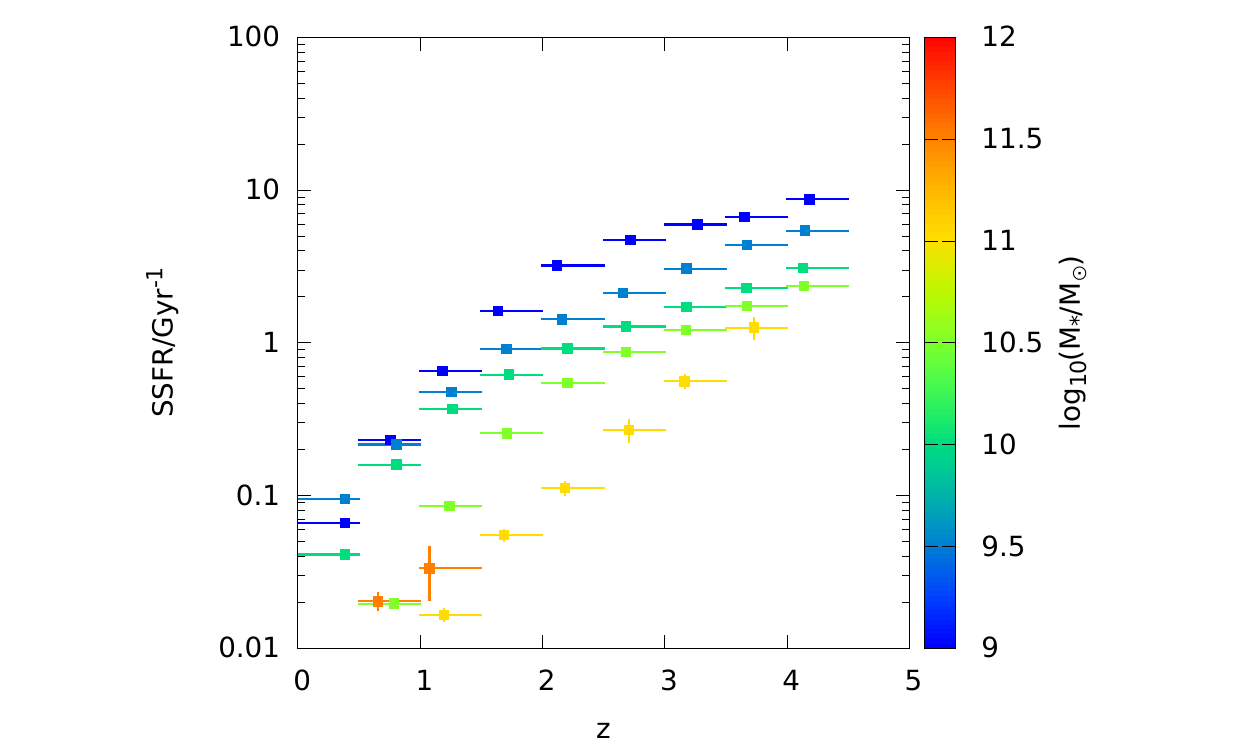}\label{fig:mill:SSFR_v_z}}
\caption{SSFR as a function of (a) stellar mass given redshift and (b)
  redshift given stellar mass for all galaxies in the
  \citet{henriques2012} simulations. \updated{The squares are centred
    on the median stellar mass in each bin.} \referee{Horizontal bars
    indicate the width of those bins, while the vertical }error
  bars simply reflect the poisson uncertainties. \referee{We have
    included the points for $3<z<5$ since they are available.}\label{fig:mill}}
\end{figure*}

Figure \ref{fig:mill} shows SSFR as a function of mass and redshift
for this simulation. Overall, SSFR amplitudes compare favourably
between the \citet{henriques2012} and VIDEO samples. It is evident
that the trend of SSFR against redshift, and against mass, is roughly
the same for the simulated and radio-derived SSFRs: SSFR increases
with redshift but decreases with stellar mass.

This tendency for SSFR to decrease with stellar mass
(Figure~\ref{fig:mill:SSFR_v_M}), has the slope $\beta$
\textit{steepening} at higher redshift, in contradiction to our
findings and those from \cite{karim2011} (see section
\ref{sec:results:comparison:karim11}). The SSFRs in the
\cite{henriques2012} simulation are lower than ours and those from
\cite{karim2011}, but on the other hand the \cite{henriques2012}
results are more consistent with those from
UKIDSS--UDS. Figure~\ref{fig:mill:SSFR_v_M} reveals that at low $z$
the SSFR range is two decades in SSFR, but this reduces to about one
decade at high $z$. For the redshift evolution of SSFR
(Figure~\ref{fig:mill:SSFR_v_z}), $n$ is very roughly constant with
stellar mass, which is consistent with our values and those of
\cite{karim2011} for all galaxies. The SSFR--$M_*$ bimodality (see
section \ref{sec:results:comparison:karim11:mstar}) is present in the
\cite{henriques2012} simulation, but only hinted at in our data at the
lowest redshifts.

\begin{figure*}
\centering
\subfigure[]{
\includegraphics[trim = 0mm 0mm 0mm 0mm, width=8cm,origin=br,angle=0]{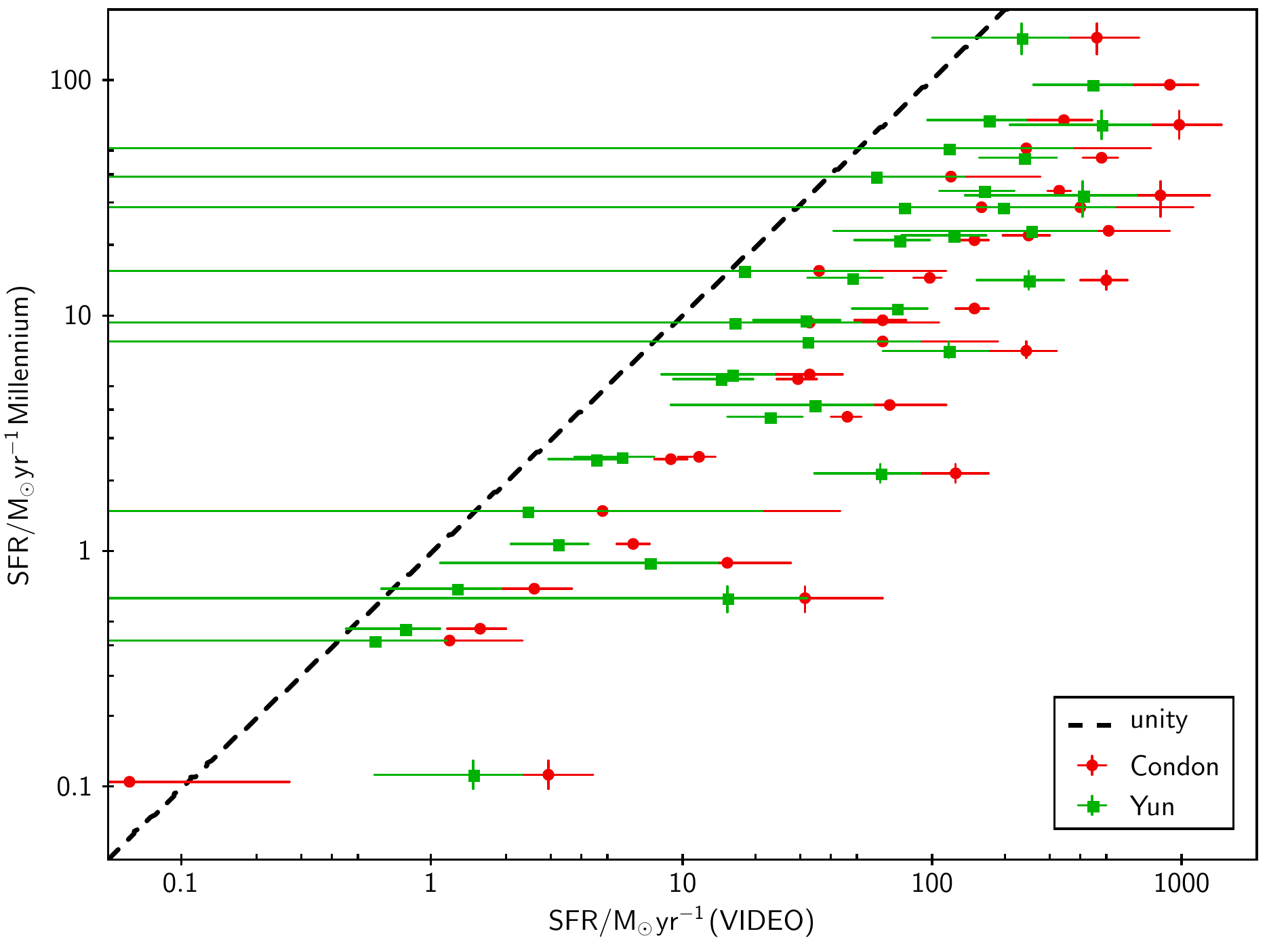}\label{fig:mill-compare:sfr}}
\subfigure[]{
\includegraphics[trim = 0mm 0mm 0mm 0mm, width=8cm,origin=br,angle=0]{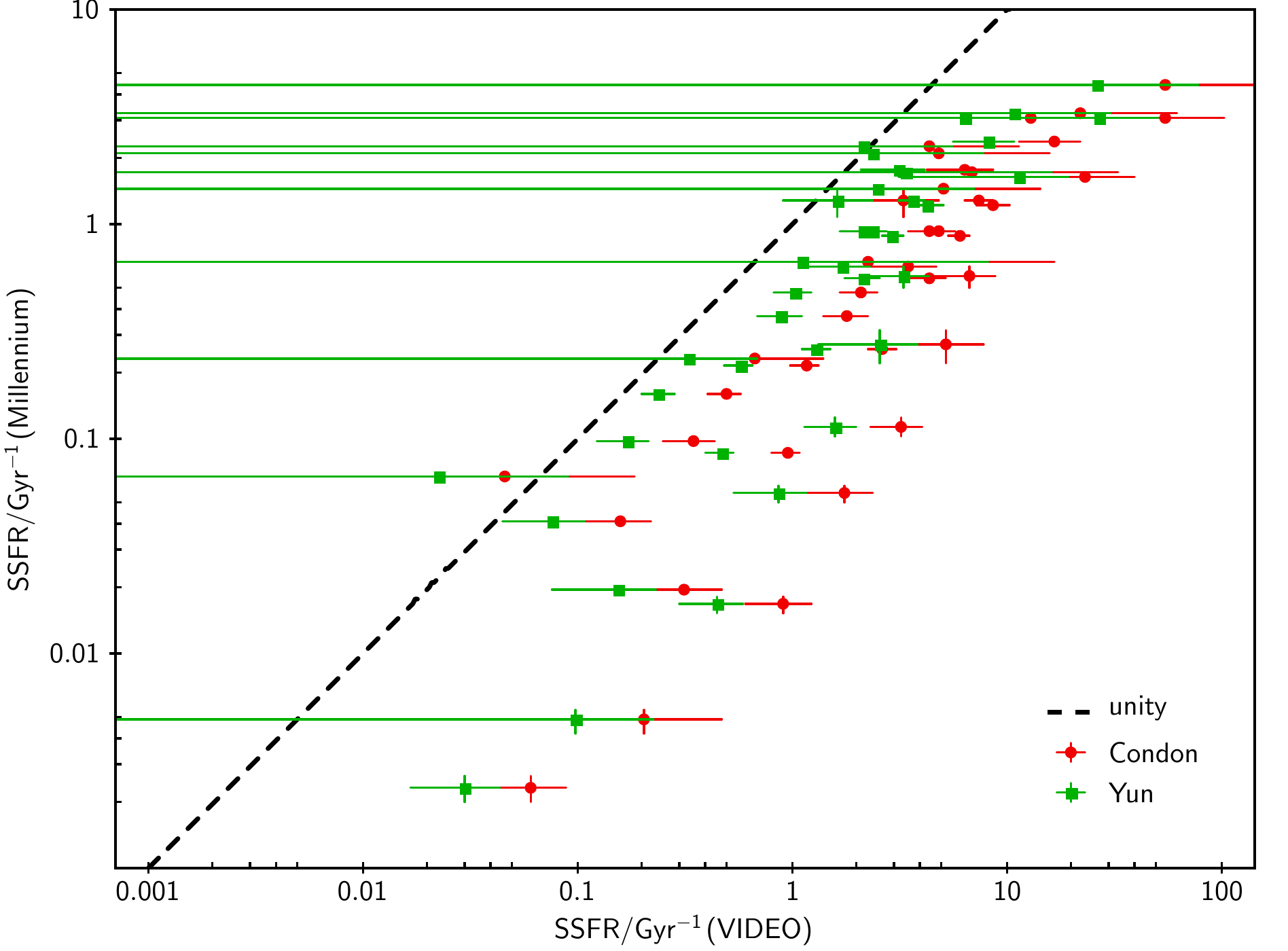}\label{fig:mill-compare:ssfr}}
\caption{Comparison of star-formation rates from this work and from
  the \citet{henriques2012} simulation: The diagrams show median (a)
  \referee{SFR} and (b) \referee{S}SFR for each mass-redshift bin
  \updated{(in this case, $0<z<5$)}. In
  each one, the \citet{condon2002} conversion (equation
  \ref{eqn:conversion:condon}) is indicated with red circles while the
  green squares use that from \citet{yun2001}, which differs solely by
  a factor of two. Unity slope is marked with a black dashed
  line.\label{fig:mill-compare}}
\end{figure*}

Directly comparing (Figure \ref{fig:mill-compare}) the star-formation
rates from this work to those from the \cite{henriques2012}
simulation, for each mass-redshift bin, the \cite{henriques2012} SFRs
and those presented here are generally inconsistent. From Figure
\ref{fig:mill-compare:sfr} we cannot tell whether the observed
difference between simulated and measured SSFRs is due to
discrepancies in SFRs or in stellar masses; Figure
\ref{fig:mill-compare:ssfr} confirms that the discrepancy could be in
(at least) the SFRs, since the VIDEO SSFRs are, like the SFRs,
consistently higher than those derived from the \cite{henriques2012}
simulation. \cite{serjeant2008} stacked SCUBA and \textit{Spitzer}
data and like us compared their results to those from the
\cite{henriques2012} simulation of \cite{delucia2006}. They also found
a large discrepancy between their submm observations and the
\cite{henriques2012} simulated SSFRs. They consider several possible
explanations for this inconsistency: (i) photometric redshift errors
(discounted by \citealt{serjeant2008}), (ii) a top-heavy initial mass
function in star-forming galaxies, or (iii) that observed submm fluxes
are controlled by cirrus heated by galaxies' interstellar radiation
fields. In our case, either the simulation may be underestimating the
SFR (as suggested by \citealt{serjeant2008}), or low-level AGN
activity may be contributing spuriously to the observed radio
SFRs. Although it is not possible to distinguish these two scenarios
without, for example, high-resolution radio imaging or X-ray data to
isolate AGN cores, we argue in section
\ref{sec:results:comparison:agn-contam} that AGN activity is not the
dominant effect.

\subsection{Far-Infrared Star-Formation Rate Indicators}
\label{sec:results:comparison:elbaz11}

Using \textit{Herschel} (\citealt{poglitsch2006, griffin2007}) data
from the Great Observatories Origins Deep Survey (GOODS;
\citealt{dickinson2003}), \cite{elbaz2011} found evolution of SSFR
with redshift for star-forming galaxies (see their Figure 18),
although it is not clear to what mass range their data are sensitive.
Their star-formation indicator is the so-called IR8, the ratio of the
24- and 8-$\mu$m luminosities. Stacking from $24\mu$m into
PACS-$100\mu$m leads to a measurement of evolution $n$ of SSFR with
redshift that is extremely similar to ours if we equate our full
sample with their `normal star-forming galaxies', the median SSFRs
both rising from $0.1\,\mathrm{Gyr}^{-1}$ at $z=0$ to $\approx
5\,\mathrm{Gyr}^{-1}$ by $z=3$.  As noted earlier, our starburst SSFRs
fall comfortably within the zone specified by \cite{elbaz2011},
i.e.,~the two data sets are compatible.

\citet{rodighiero2010} derived SSFRs directly from 24-$\mu$m
\textit{Herschel}-PACS over GOODS--N, finding instead for this
star-forming sample that although SSFR does increase with redshift
(for $M_*>10^{11}M_{\odot}$) by a factor of about 15 from $z=0$ to
$z=2$, it has already flattened off by $z\approx 1.5$. In their
analysis for a star-forming sample, the general trend of downsizing is
once again upheld; however, the dependence of SSFR on stellar mass
\textit{steepens} with redshift ($\mathrm{d}\beta/\mathrm{d}z<0$),
from a flat slope ($\beta=0$) at $z<1$ to
$\beta=-0.50^{+0.13}_{-0.16}$ by $z\approx 2$, in broad agreement with
the \cite{dunne2009} result. This is still consistent with our (and
the \citealt{karim2011}) scenario in which $\beta_{sbn}$ is
independent of $z$.

SFRs were measured by \cite{wang2013} for 3.6-$\mu$m-selected
star-forming galaxies in the \textit{Herschel} Multi-tiered
Extragalactic Survey (HerMES; \citealt{hermes}). One of the three
fields studied is in fact the COSMOS field with the \cite{karim2011}
photometric redshifts and stellar masses determined using \textsc{Le
  Phare}. Rather than extrapolating the infrared luminosity
$L_{\mathrm{IR}}$ (8--1000\,$\mu$m) from 24-$\mu$m data, SEDs are
fitted at 24\,$\mu$m and to the FIR bands of 250, 350 and 500\,$\mu$m
where possible.  $L_{\mathrm{IR}}$ in turn gives the SFR via the
simple \cite{kennicutt1998a} relation. For undetected (at SPIRE
wavelength) galaxies, which make up 70~per~cent of the sample, the SFR
is calculated from the optical SED fit. \cite{wang2013}'s SSFR--$M_*$
slope $\beta$ flattens slightly with redshift
($\mathrm{d}\beta/\mathrm{d}z>0$, up to $z=2$), but the gradient is
still compatible with our measured independence of $\beta$ with $z$.

In summary, downsizing prevails irrespective of the waveband used. A
key result is that there is broad agreement between indicators
(including ours) that SSFR flattens off by $z\approx3$. Although the
level of the SSFR--mass slope $\beta\approx -0.5$ is roughly
consistent for star-forming samples across indicators, there is not a
clear picture on its evolution with redshift $\beta(z)$, and more work
is needed in this area.

\subsection{AGN Contamination}
\label{sec:results:comparison:agn-contam}

AGN contamination could be a major reason to doubt that 1.4-GHz radio
emission might reliably trace star formation. \cite{dunne2009}
demonstrate that there are two reasons why this is not a serious
concern:

\begin{enumerate}
\item First, \citet{reddy2005} and \citet{daddi2007} both measure
  contamination to SFRs from X-ray-emitting AGN, but find that
  contamination due to such sources decreases significantly as one
  moves to fainter $K$-band fluxes (just 3--4 per cent by $K$=22.9,
  down from up to $\approx 30$ per cent at $K$=19.9). There is also
  circumstantial evidence (supported by \citealt{fomalont2006},
  \citealt{bondi2007} and \citealt{simpson2012}) that the AGN fraction
  decreases significantly below 1.4-GHz fluxes of $\approx
  100\,\mu$Jy.
\item Second, radio-loud Seyfert galaxies may cause a departure from
  the radio-FIR correlation, but \cite{dunne2009} demonstrate that
  this is not the case by comparing their radio-derived results to
  those from the submm.
\end{enumerate}

\cite{muxlow2005} made deep ($\approx 3\,\mu$Jy) 1.4-GHz VLA images of
the \textit{Hubble} Deep Field, determining that the proportion of
starburst systems increases with decreasing flux right down to that
level. Nearly all their faintest detections are resolved and so
unlikely to be AGN. This was also predicted in the SKADS simulations
of \cite{wilman2008,wilman2010}.

\cite{karim2011} further found no evidence for AGN contamination in
their analysis, and point out that median stacking would tolerate such
contamination even if it were present. This highlights the resistance
of our median-stacking results to moderate levels of AGN
contamination. \cite{pannella2009} took particular care to clean AGN
from their sBzK sample; any contamination in our VLA data would tend
to flatten $\beta$, but if anything we see lower/consistent values
than their already flat value of $\beta$. We therefore argue that AGN
contamination is not a significant effect for our data set.

\section{Conclusions}
\label{sec:conclusions}

We have stacked deep ($17.5\,\mu$Jy) VLA radio observations at the
positions of \Ks-selected sources in the VIDEO field (for \Ks$<$23.5,
sensitive to $0<z\lesssim5$). Stellar masses and redshifts were
estimated by fitting spectral templates to 10-band photometry. We
separated galaxies into different populations (ellipticals, irregulars
and starbursts) based on their best-fitting spectral
classification. We used median single-pixel stacking, converting the
stacked radio fluxes to star-formation rates using the
\citet{condon2002} relation. Specific star-formation rates, as
expected, were highest for starburst galaxies and lowest for
ellipticals.

\begin{enumerate}

\item We subdivided the samples into stellar-mass and redshift bins,
  then fitted specific star-formation rates as a separable function of
  stellar mass and redshift in each bin. We found that SSFR falls with
  stellar mass for both our full and starburst samples. Hence the
  `downsizing' scenario is supported by our data because we measure
  $\beta<0$.

\item The SSFR--mass gradients $\beta$ became less steep with redshift
  (from $\beta\approx -0.75$ to $\beta\approx -0.25$ out to
  $z\simeq2$) for the full and elliptical samples, but were
  independent of redshift ($\beta\approx -0.5$) for the starburst and
  irregular galaxies.

\item We have compared our results to those from other radio
  star-formation rate indicators. We found that the evolution of the
  SSFR--mass slopes as a function of redshift, for both our full and
  starburst samples, are especially consistent with those from the
  COSMOS study by \cite{karim2011}, becoming less steep out to
  $z\simeq2$ for the full sample. Given methodological differences,
  our results are also consistent with those of \cite{dunne2009}. A
  bimodality in the SSFR--mass plane present in the observational work
  of \cite{dunne2009} and \cite{karim2011} and in the semi-analytic
  models of \cite{henriques2012} is only hinted at in our two lowest
  redshift bins.

\item We compared our results to those generated from the
  \cite{henriques2012} simulation, and discovered (in all cases for
  our full sample) that the SSFR--mass slopes are highly inconsistent
  with those from our study and from \cite{karim2011}, steepening with
  redshift where ours both flatten out.

\item For far-infrared indicators covering star-forming samples, the
  picture that emerges from the literature is not a clear one, but our
  result that $\beta$ is independent of $z$ lies somewhere between the
  extreme cases of \cite{rodighiero2010} and \cite{wang2013}.

\item For the SSFR--redshift relation, we found evolution to be
  fastest among lower-mass ellipticals, whereas starbursts and
  irregulars tend to co-evolve independent of mass. \referee{The rate
    of evolution reduces} as higher-redshift objects are included.

\item On the topic of AGN contamination, there is strong evidence from
  the literature that this would not adversely affect out
  results. However, we do note that high-resolution radio (e.g.~from
  the forthcoming MeerKAT telescope, e-MERLIN or VLBI) or X-ray
  imaging (cf.~\citealt{pannella2009}) would be beneficial in
  resolving this issue.

\end{enumerate}

%======================================================
\section*{Acknowledgments}
\label{ack}

JZ and RD gratefully acknowledge South Africa National Research
Foundation Square Kilometre Array Research Fellowships. \updated{MJ,
  KK and NM are grateful to the South Africa National Research
  Foundation Square Kilometre Array Project for financial support.} We
thank Russell Johnston, Mathew Smith, Matthew Prescott, Mattia
Vaccari, Lingyu Wang, Bruce Bassett and Michelle Knights for useful
discussions. \updated{We are thankful to the anonymous referee for
  their helpful comments.} The project was initiated at the Cape Town
International Cosmology School in January 2012, organized by AIMS,
ICTP and UWC, and continued at the PARSLEY 2013 workshop of the
University of the Western Cape.

%======================================================
\bibliographystyle{mn2e}
\bibliography{stacking}\label{lastpage}

%%%%%%%%%%%%%%%%%%%%%%%%%%%%%%%%%%%%%%%%%%%%%%%%%%%%%%%%%%%%%

\bsp

\end{document}